\documentclass[aps,prl,reprint,superscriptaddress,amsmath,amssymb,floatfix]{revtex4-2}

\usepackage{graphicx}
\usepackage{array}
\usepackage{bm}
\usepackage{xcolor}
\usepackage[colorlinks=true,linkcolor=blue!60!black,citecolor=blue!60!black,urlcolor=blue!60!black]{hyperref}

\newcommand{\LE}{\mathcal{L}_E}
\newcommand{\LG}{\mathcal{L}_G}
\newcommand{\LD}{\mathcal{L}_D}
\newcommand{\xc}{\mathrm{xc}}
\newcommand{\refr}{\mathrm{ref}}
\newcommand{\scf}{\mathrm{SCF}}
\newcommand{\phib}{\bar\phi}
\newcommand{\kcal}{kcal/mol}

\begin{document}

\title{Fitting Conventional and Neural Exchange-Correlation Functionals\\ to Accurate Densities without Self-Consistent Fields}

\author{Yuhao Chen}
\affiliation{Shanghai Engineering Research Center of Molecular Therapeutics and New Drug Development, Shanghai Frontiers Science Center of Molecule Intelligent Syntheses, School of Chemistry and Molecular Engineering, East China Normal University, Shanghai 200062, China}

\author{Donald G. Truhlar}
\affiliation{Department of Chemistry, Chemical Theory Center, and Minnesota Supercomputing Institute, University of Minnesota, Minneapolis, Minnesota 55455-0431, USA}

\author{Xiao He}
\email{xiaohe@phy.ecnu.edu.cn}
\affiliation{Shanghai Engineering Research Center of Molecular Therapeutics and New Drug Development, Shanghai Frontiers Science Center of Molecule Intelligent Syntheses, School of Chemistry and Molecular Engineering, East China Normal University, Shanghai 200062, China}
\affiliation{Anhui Provincial Key Laboratory of Advanced Catalysis and Energy Materials, Anhui Ultra High Molecular Weight Polyethylene Fiber Engineering Research Center, School of Chemistry and Chemical Engineering, Anqing Normal University, Anqing 261433, China}
\affiliation{New York University--East China Normal University Center for Computational Chemistry, New York University Shanghai, Shanghai 200062, China}

% PRL limit: 600 characters including spaces. Current: 597.
\begin{abstract}
Exchange-correlation functionals are fit on fixed orbitals but used self-consistently, so better energies often bring worse densities. We introduce a low-cost workflow that predicts in closed form, from one Fock build on inverse Kohn-Sham orbitals of any accurate density, a functional's self-consistent density, and fits it to that density with no self-consistent-field solution. For the MN15 form the fit is convex; the refitted hybrid beats MN15 on unseen energies, electric-field properties, and densities (29-fold). A neural meta-GGA fitted to 6 energies and 11 densities beats SCAN in both.
\end{abstract}

\maketitle

\begin{figure*}[t]
\includegraphics[width=\textwidth]{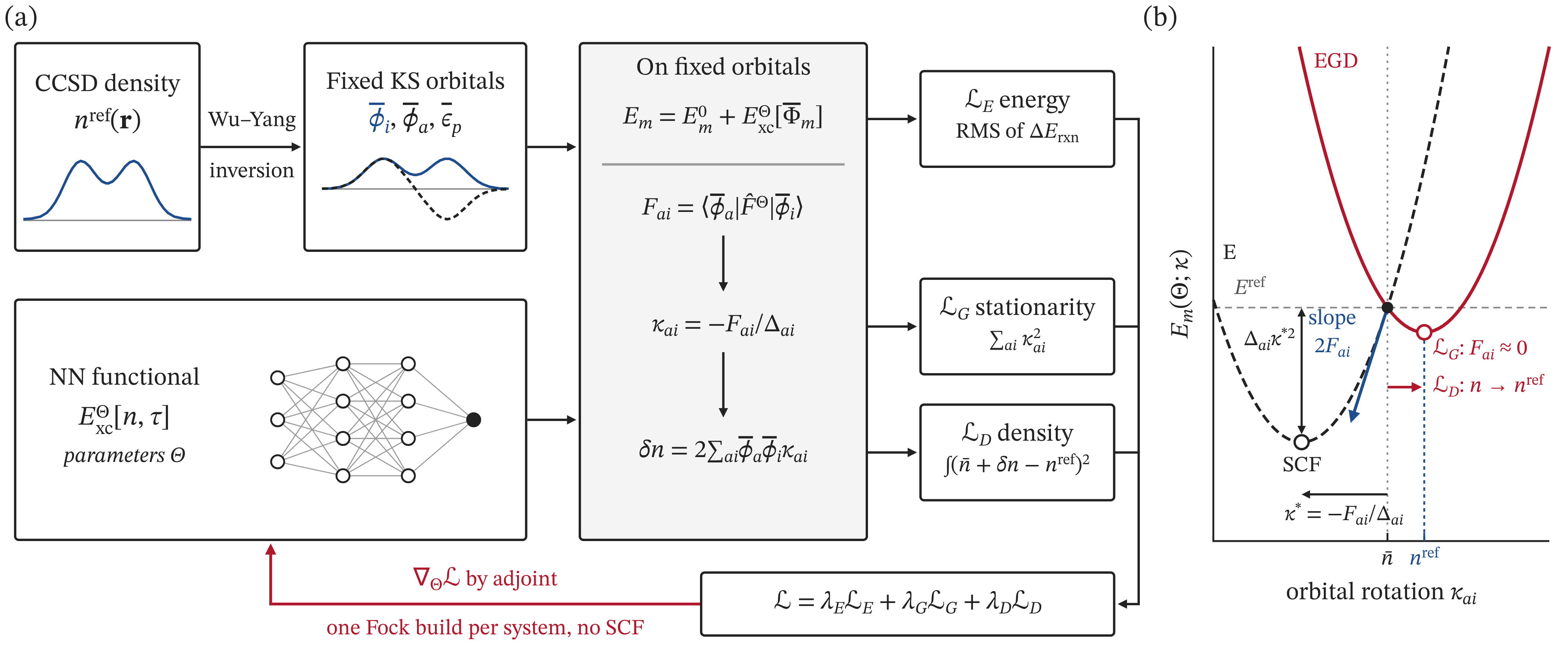}
\caption{Training XC functionals on fixed orbitals.
(a)~Wu-Yang inversion of an accurate reference density (here CCSD) gives fixed KS orbitals, on which the occupied-virtual Fock block $F_{ai}$ of the trained functional yields the predicted SCF rotation $\kappa_{ai}$ (stationarity loss $\LG$) and the first-order density change $\delta n$ (density loss $\LD$); the energy loss $\LE$ uses the same orbitals. One Fock build per system gives the parameter gradient, with no SCF.
(b)~Energy along one occupied-virtual rotation. Energy-only training (E) matches the reference energy at $\kappa=0$, but the orbital gradient $2F_{ai}$ drives the SCF solution to $\kappa^{*}=-F_{ai}/\Delta_{ai}$, causing an energy error $\Delta_{ai}\kappa^{*2}$ and a density shift $\delta n$. EGD makes the reference orbitals stationary.}
\label{fig:scheme}
\end{figure*}

% =====================================================================
%  INTRODUCTION:  stakes -> mechanism (gap) -> prior routes -> this Letter
% =====================================================================
Kohn-Sham density functional theory (KS-DFT)~\cite{Hohenberg1964,Kohn1965} is the most widely used electronic-structure method in chemistry and materials science, and its accuracy is set by the approximation to the exchange-correlation (XC) energy. The most accurate approximations for molecular thermochemistry contain tens of empirical parameters~\cite{Peverati2014,Mardirossian2016,Yu2016MN15,Mardirossian2017} or, as machine-learned functionals, thousands or more~\cite{Nagai2020,Dick2020,Kirkpatrick2021,Bystrom2022,Pederson2022,Luise2025,Liang2026COACH}. These parameters are fixed almost exclusively by reference energies; the electron density does not enter the fit. Medvedev \textit{et al.}\ found that functionals with more empirical parameters tend to give worse atomic densities~\cite{Medvedev2017}, suggesting that part of the energetic gain comes from error cancellation rather than from approaching the exact functional~\cite{Medvedev2017,Kepp2017,Brorsen2017}.

A concrete but rarely discussed mechanism underlies this trend: a functional is not used in the state in which it is fitted. Its parameters are fitted to energies on fixed orbitals, but the functional is solved self-consistently. A parameter change that lowers the fixed-orbital energy error also changes the XC potential, so the self-consistent-field (SCF) solution relaxes away from the fitted orbitals---and an energy-only loss does not penalize this relaxation, even when the fixed orbitals come from the exact density. In the language of density-corrected DFT~\cite{Kim2013,Wasserman2017}, energy fitting controls the functional-driven error but leaves the density-driven error unconstrained. Accuracy reached in the fit therefore need not survive self-consistency, let alone transfer to new molecules.

Two routes bring information beyond energies into the fit. One fits XC potentials obtained by density inversion~\cite{Zhao1994,Wu2003}, as in the TH, HCTH, and B97-2 functionals~\cite{Tozer1996,Tozer1998,Handy1998,Hamprecht1998,Wilson2001}, or learns the density-to-potential map~\cite{Nagai2018,Zhou2019}; but finite-basis inversion is ill-conditioned and fixes the potential only up to a constant~\cite{Staroverov2006,Jensen2018,Kanungo2019,Shi2021}, so pointwise fitting imports numerical noise into the target. The other is differentiable DFT~\cite{Li2021,Kasim2021,Dick2021}, which places the SCF cycle inside automatic differentiation, so that every training step must converge the SCF and backpropagate through it, which is costly and fragile for hard-to-converge systems; alternating training with SCF~\cite{Chen2021} likewise requires repeated SCF solutions. Residual penalties have also pushed fixed-density training toward self-consistency~\cite{Kirkpatrick2021,delMazo2023}, but they make the functional stationary at orbitals of approximate functionals, which is not self-consistency at the correct density, and the density itself never enters the loss. Because every step that uses densities must converge an SCF, density-informed fits have stayed small: the energies and densities of three molecules~\cite{Nagai2020}, six energies and nine density profiles~\cite{Kasim2021}, or 21 molecules with 26 energies and coupled-cluster densities~\cite{Dick2021}. Accurate densities therefore have yet to shape a functional fitted to the broad energy databases that define modern thermochemistry.

In this Letter we introduce a fitting workflow that makes an accurate density the direct target of the self-consistent solution, without solving the SCF (Fig.~\ref{fig:scheme}). Its key idea is that the density a trained functional would converge to can be predicted in closed form: on the inverse-KS orbitals of an accurate reference density, the Brillouin residual of the functional---the occupied-virtual block of its Fock matrix---fixes, to first order, the SCF orbital rotation, the SCF density shift, and the gap between fitted and self-consistent energies. Three elements follow. (i)~A \emph{first-order density} loss compares this predicted self-consistent density with the reference density, so the observable itself is fitted from one Fock build per system. (ii)~Anchoring the reference to inverse-KS orbitals of an accurate density, from any correlated method, makes the target exact self-consistency at that density and admits only the potential error that rotates occupied orbitals; the constant and ill-determined components that plague inversion drop out. (iii)~A \emph{stationarity} loss on the predicted rotation, weighted by the inverse orbital Hessian, keeps the first-order prediction valid and retains its pull near stationarity. For linear functionals the whole fit is convex, and the main payoff is for conventional functionals: refitting the MN15 form~\cite{Yu2016MN15} to 1500 reaction energies under the density constraint yields a hybrid that improves on MN15 for held-out energies and unfitted electric-field properties and, on 228 unseen density records, has the lowest density error of eight hybrids, 29 times below that of MN15. With 1500 energies and 98 density molecules, this fit uses more than 50 times the energy data and more than 4 times the density data of the largest earlier density-informed fit~\cite{Dick2021}, and it is tested on 228 further densities. The same workflow trains a neural meta-GGA from 6 energies and the densities of 11 small systems that reverses the trend of Ref.~\cite{Medvedev2017} seen under energy-only training and surpasses SCAN in energies and densities.

% =====================================================================
%  THEORY
% =====================================================================
\begin{figure*}[t]
\includegraphics[width=\textwidth]{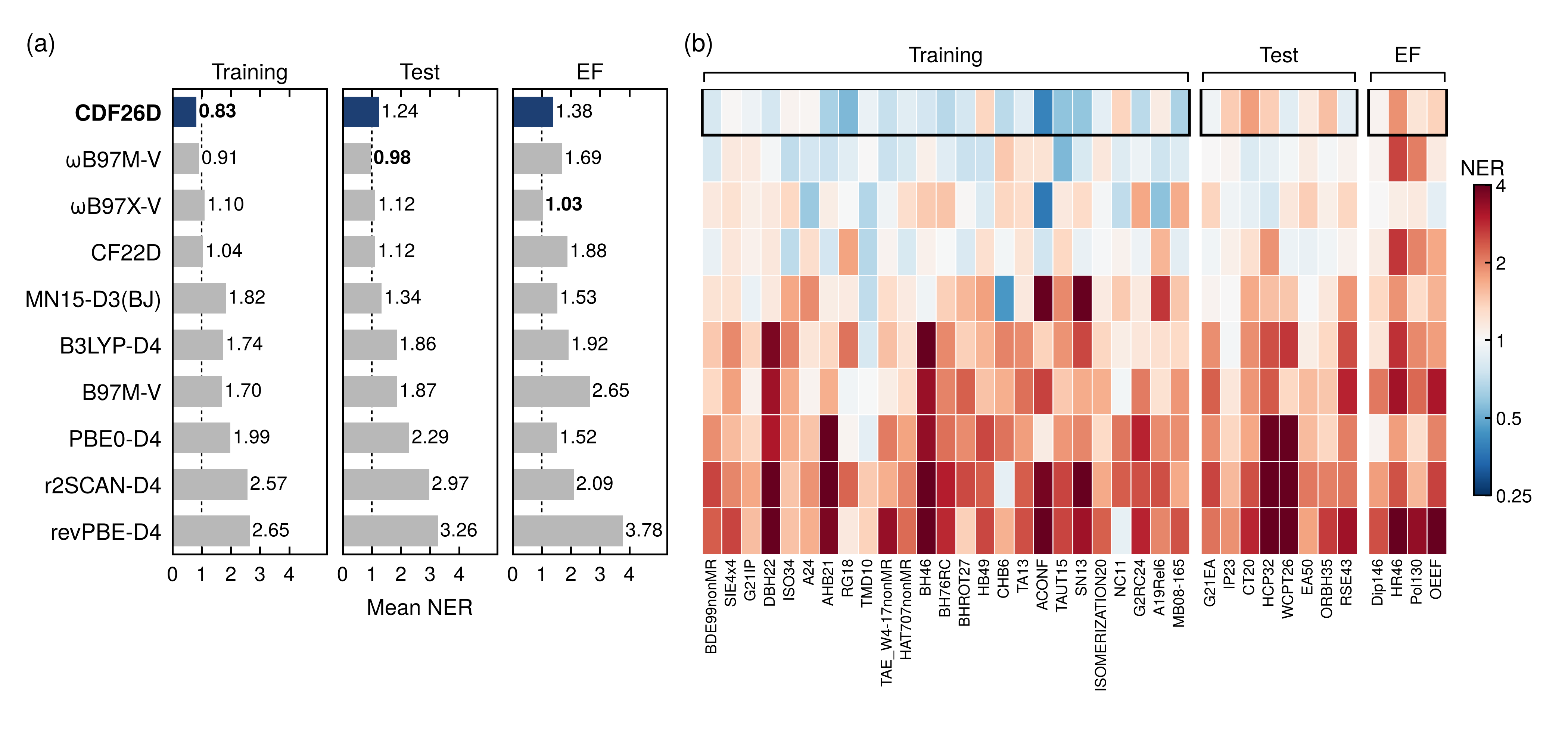}
\caption{Self-consistent normalized error ratios (NER) of CDF26D and representative functionals on GSCDB137~\cite{Liang2025GSCDB}; NER is the error relative to the mean of the second- to fourth-lowest hybrid errors for each set, and NER$\,<1$ beats leading hybrids. Training: the 25 fitted sets (1500 data); Test: 8 sets not used in fitting (254 data); EF: electric-field sets, not fitted. (a)~Mean NER over each group. (b)~NER of each set; rows follow the order in (a), CDF26D outlined. Comparison functionals: $\omega$B97M-V~\cite{Mardirossian2016}, $\omega$B97X-V~\cite{Mardirossian2014}, CF22D~\cite{Liu2023CF22D}, MN15-D3(BJ)~\cite{Yu2016MN15}, and others in the SM~\cite{SM}.}
\label{fig:cdf}
\end{figure*}

\textit{Fixed-orbital fitting and its self-consistent consequences.}---Let $\{\phib_{p\sigma},\bar\epsilon_{p\sigma}\}$ be the KS orbitals and orbital energies obtained from an accurate reference density $n^\refr$ (CCSD in this work) by regularized Wu-Yang inversion~\cite{Wu2003}, i.e., eigenstates of a local-potential Hamiltonian $\hat h^\refr_\sigma$ (SM Secs.~S2 and S3~\cite{SM}). On these fixed orbitals the energy of system $m$ is $E_m(\Theta)=E^0_m+E^\Theta_\xc[\bar\Phi_m]$, where $E^0_m$ does not depend on the parameters $\Theta$; energy-only fitting (E) matches reaction energies assembled from $E_m(\Theta)$ to reference values. Rotating the occupied orbitals, $\phi_{i\sigma}=\phib_{i\sigma}+\sum_a\kappa^\sigma_{ai}\phib_{a\sigma}$, and expanding to second order gives
\begin{equation}
E_m(\Theta;\kappa)\simeq E_m(\Theta)+\sum_{\sigma ai}\big[2F^\sigma_{ai}\kappa^\sigma_{ai}+\Delta^\sigma_{ai}(\kappa^\sigma_{ai})^2\big],
\label{eq:expansion}
\end{equation}
where $F^\sigma_{ai}(\Theta)=\langle\phib_{a\sigma}|\hat F^\sigma_\Theta|\phib_{i\sigma}\rangle$, the Brillouin residual, is the occupied-virtual block of the trained functional's (generalized) Fock operator on the reference orbitals, and $\Delta^\sigma_{ai}=\max(\bar\epsilon_{a\sigma}-\bar\epsilon_{i\sigma},0.05\,E_h)$ is the uncoupled orbital Hessian. Its minimum gives the predicted SCF rotation and its two observable consequences,
\begin{align}
\kappa^\sigma_{ai}&=-F^\sigma_{ai}/\Delta^\sigma_{ai},\label{eq:kappa}\\
E^\scf_m&\simeq E_m-\textstyle\sum_{\sigma ai}\Delta^\sigma_{ai}(\kappa^\sigma_{ai})^2,\label{eq:Escf}\\
\delta n_m(\mathbf r)&\simeq 2\textstyle\sum_{\sigma ai}\phib_{a\sigma}(\mathbf r)\phib_{i\sigma}(\mathbf r)\kappa^\sigma_{ai}.\label{eq:dn}
\end{align}
Equation~\eqref{eq:Escf} is the density-driven error~\cite{Kim2013,Wasserman2017} that E fitting leaves unconstrained, and Eq.~\eqref{eq:dn} predicts in closed form the density the functional would reach self-consistently. Both vanish if and only if $F^\sigma_{ai}=0$, i.e., if the reference orbitals are a self-consistent solution of the trained functional in the given basis~\cite{Helgaker2000}; the uncoupled Hessian only weights deviations from this exact target, and a positive Hartree-XC kernel screens the true relaxation (SM Sec.~S6). Because $\hat h^\refr$ is diagonal in the reference orbitals, $F^\sigma_{ai}=\langle\phib_{a\sigma}|\hat v^\Theta_{\xc,\sigma}-v^\refr_{\xc,\sigma}|\phib_{i\sigma}\rangle$, so only the occupied-virtual block of the potential error enters: a constant shift cancels, and the components with a vanishing occupied-virtual block are exactly the null space of the inversion Hessian (SM Eq.~S28), the ill-determined part that limits pointwise potential fitting. For meta-GGAs and hybrids, $\hat v^\Theta_\xc$ includes the nonmultiplicative $\tau$ term~\cite{Seidl1996,Neumann1996} and nonlocal exact exchange.

\textit{Losses.}---The first-order density loss, the central element of the workflow, compares the predicted self-consistent density with the reference density,
\begin{equation}
\LD=\frac1{N_{\mathcal S}}\sum_{m\in\mathcal S}\int\big[\bar n_m(\mathbf r)+\delta n_m(\mathbf r)-n^\refr_m(\mathbf r)\big]^2d\mathbf r,
\label{eq:LD}
\end{equation}
where $\bar n_m$ is the density of the reference orbitals, so $\LD$ also absorbs the representable part of the finite inversion residual. The stationarity loss $\LG=\sum_{m\in\mathcal S}\sum_{\sigma ai}(\kappa^\sigma_{ai})^2$ bounds the energy relaxation of Eq.~\eqref{eq:Escf} by $\max\Delta\cdot\LG$. Unlike the squared energy relaxation ($\propto F^4$)~\cite{Kirkpatrick2021} or the unweighted residual norm~\cite{delMazo2023}, it measures the orbital change directly, keeps a finite gradient near stationarity, weights small-gap pairs, and keeps $\LD$ in its linear regime. The objective is $\mathcal L=\lambda_E\LE+\lambda_G\LG+\lambda_D\LD$, with $\LE$ the root-mean-square error of the reference reaction energies on fixed orbitals. Each update needs one Fock build per system and one adjoint pass for $\nabla_\Theta\mathcal L$, with no SCF solved or differentiated (SM Secs.~S7--S11); every energy and density reported below is fully self-consistent.

% =====================================================================
%  RESULTS I: CONVENTIONAL HYBRID
% =====================================================================

\textit{Conventional hybrid functionals.}---The workflow is most effective for conventional functionals. Their parameters enter linearly, so the density constraint keeps the fit convex, and its cost of one Fock build per density system lets it scale to large training sets. We refit the 59 linear parameters of the MN15 form~\cite{Yu2016MN15} (including the exact-exchange fraction) to 1500 reaction energies in 25 data sets of GSCDB137~\cite{Liang2025GSCDB}, subject to the smoothness bound of CF22D~\cite{Liu2023CF22D}; the density information enters through the first-order density constraint alone, $\LD\le3\times10^{-4}$ on DP99, 99 reference density records (CCSD/aug-cc-pVTZ) of 98 molecules. Density errors are reported as ${\rm DP}=\int(n^\scf-n^\refr)^2d\mathbf r$ averaged over records, in units of $10^{-3}$ a.u. (SM Sec.~S14). For a linear functional, $\delta n$ is affine in the parameters, so $\LD$ is a convex quadratic and the fit is a least-squares problem with convex constraints whose Karush-Kuhn-Tucker conditions certify the global optimum, with no learning rate or loss weight to tune (SM Sec.~S13). Between fits, the energy matrices are rebuilt on the current self-consistent orbitals, as in conventional fitting, while the density constraint stays on the reference orbitals; adding data only adds matrix rows. The resulting functional, CDF26D, has 48.9\% exact exchange and D4 dispersion~\cite{Caldeweyher2019} fitted afterwards to the same sets. It keeps the MN15 form, so it costs no more than MN15 in any calculation.

All values in Fig.~\ref{fig:cdf} are self-consistent. On the training sets, CDF26D has the lowest mean NER of the 15 hybrids in GSCDB137 and this work, and on the held-out sets it improves on MN15, whose form it shares (1.24 vs 1.34). The largest gain appears where the density response matters: for electric-field properties, which enter neither the energy nor the density term of the fit, the mean NER of CDF26D is 10\% below that of MN15 and 27\% below that of CF22D, which shares the form but was fitted to 6572 energies without a density constraint~\cite{Liu2023CF22D}. Dipole moments and polarizabilities are moments of the density and of its response to a field, so a functional whose self-consistent density follows the reference should improve them without being fitted to them; the EF gain is the expected signature of this mechanism. On DP99, the constraint set, the self-consistent density error is 0.17, below the bound of 0.3 imposed on the first-order prediction; the closed-form prediction is thus conservative, as expected from the screening noted after Eq.~\eqref{eq:Escf}.

\begin{table}[t]
\caption{Self-consistent density errors, ${\rm DP}\times10^{3}$ (a.u.), of hybrid functionals on DP99 and DP228. DP99 is the constraint set of the CDF26D fit ($^\dagger$); DP228 is disjoint from it and enters no fit. Best DP228 value in bold.}
\label{tab:dp}
\begin{ruledtabular}
\begin{tabular}{lcc}
Functional & DP99 & DP228 \\
\hline
CDF26D (this work) & 0.17$^\dagger$ & \textbf{1.03} \\
$\omega$B97M-V~\cite{Mardirossian2016} & 0.40 & 1.18 \\
SCAN0~\cite{Hui2016SCAN0} & 0.34 & 1.19 \\
PBE0~\cite{Adamo1999} & 0.59 & 1.20 \\
B3LYP~\cite{Becke1993,Stephens1994} & 1.35 & 1.71 \\
M06-2X~\cite{Zhao2008M06} & 0.33 & 2.31 \\
CF22D~\cite{Liu2023CF22D} & 1.37 & 8.85 \\
MN15~\cite{Yu2016MN15} & 10.14 & 29.73 \\
\end{tabular}
\end{ruledtabular}
\end{table}

The decisive test is DP228, 228 density records disjoint from DP99 and not used in fitting (SM Sec.~S14). CDF26D has the lowest DP228 error of the eight hybrids in Table~\ref{tab:dp}: once $\LD$ enters the fit, the same functional form yields densities 29 and 8.6 times more accurate than MN15 and CF22D on molecules it has never seen. CF22D shares the form and the smoothness restraint and was fitted to four times as many energies, so this gain does not come from more energy data; it comes with the density constraint.

\textit{Why the fit scales.}---The size of the convex problem is set by the functional, not by the data. Each energy record adds one row to $B$ in Eq.~(S82) of the SM, and each density record adds one Fock build and its rows to the response matrix, which a QR factorization folds into a single $60\times60$ quadratic form however many records it contains (SM Sec.~S13). Energy and density data can therefore grow independently and cheaply: the 98 density molecules used here cost 98 fixed-orbital Fock builds, whereas a differentiable-SCF fit would converge and backpropagate 98 SCF cycles at every step~\cite{Kasim2021,Dick2021}. The same construction leaves room for densities from other correlated methods and for system classes that are hard to converge, such as transition-metal compounds.

% =====================================================================
%  RESULTS II: NEURAL FUNCTIONALS
% =====================================================================
\begin{figure*}[t]
\includegraphics[width=\textwidth]{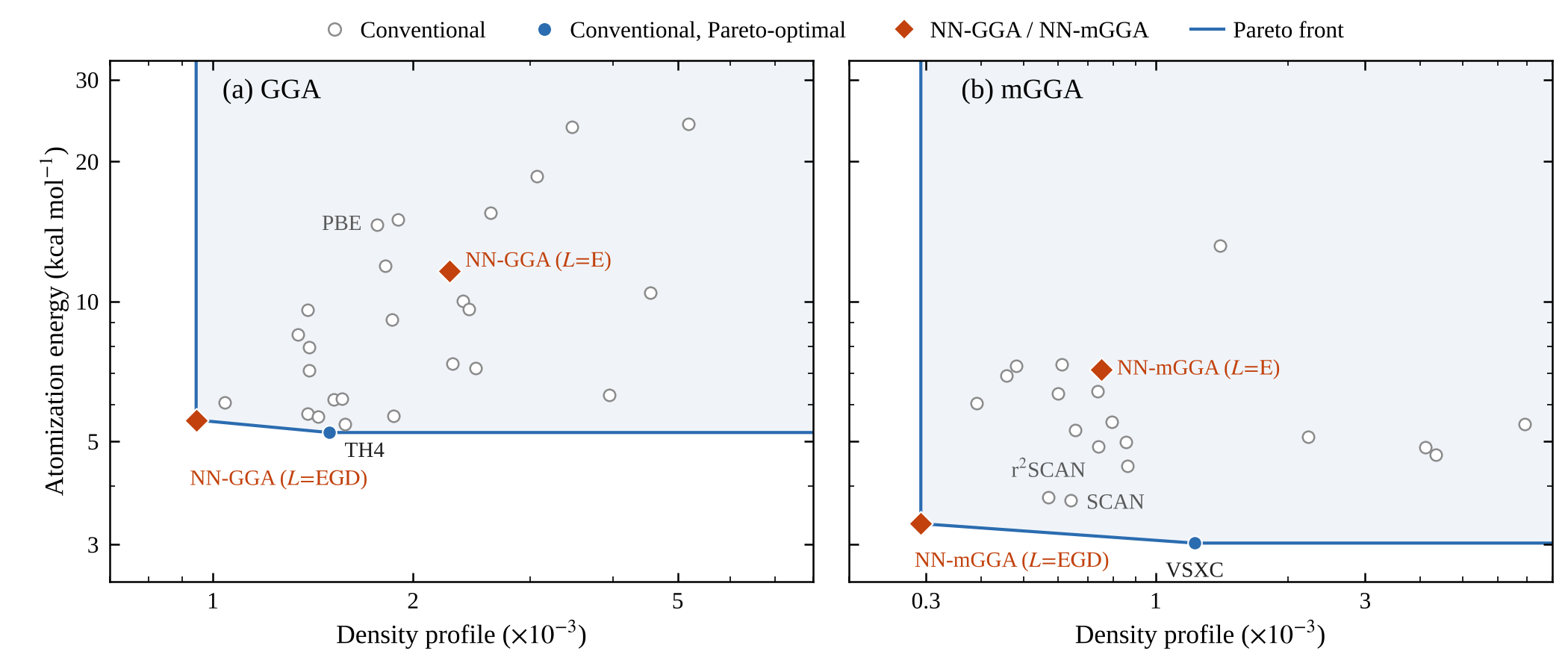}
\caption{(a)~GGA and (b)~meta-GGA functionals in the plane of AE103 atomization-energy MAE and DP99 density error. Open circles: conventional functionals; filled blue circles: Pareto-optimal conventional functionals; blue line: Pareto front; red diamonds: NN functionals trained with $L=E$ or $L=EGD$. Values for all functionals are in the SM~\cite{SM}.}
\label{fig:pareto}
\end{figure*}

\textit{Neural functionals from minimal data.}---The same workflow trains neural functionals from very little data. We correct PBE and SCAN~\cite{Perdew1996,Sun2015} as $E^\Theta_\xc=\alpha E^{\rm base}_\xc+\beta\int n\,f_\theta(\mathbf z)\,d\mathbf r$, where $f_\theta$ is a network of $\ln(1+n)$, the spin polarization, $\ln(1+s)$, and, for the meta-GGA, $\ln(1+a_\tau)$ (2274 and 2306 parameters; SM Sec.~S4). Training uses four W4-11 atomization energies~\cite{Karton2011}, two experimental ionization energies~\cite{NISTASD}, and reference densities (UCCSD $\Lambda$~\cite{Purvis1982,Salter1989}) of 11 atoms, ions, and diatomics (aug-cc-pVTZ~\cite{Dunning1989,Kendall1992}; split of Ref.~\cite{Kasim2021} with O$^+$ and Ne$^+$ added, SM Table~S1). We compare energy-only training (E, $\lambda_G=\lambda_D=0$) with the full objective (EGD; SM Sec.~S11) on the self-consistent atomization energies of 103 molecules (AE103)~\cite{Kasim2021}, many with elements and bonds absent from training, and on DP99, which enters no neural fit. These are essentially the training data of the differentiable-DFT functional of Ref.~\cite{Kasim2021}, here fitted with no SCF at any step.

Energy-only training reproduces the trend of Ref.~\cite{Medvedev2017}: both E models have larger density errors than their base functionals (Fig.~\ref{fig:pareto}; NN-GGA 2.3 vs PBE 1.8, NN-mGGA 0.75 vs SCAN 0.64). EGD reverses it and moves both networks to the low-density-error end of the Pareto front. NN-GGA(EGD) lowers the AE103 mean absolute error from 14.6 (PBE) to 5.6~\kcal\ with the lowest density error of all GGAs examined, 0.94, reaching the end of the GGA front previously held by TH3 and TH4, which were fitted to ab initio XC potentials~\cite{Handy1998}. NN-mGGA(EGD) reaches 3.3~\kcal\ and 0.29, beating SCAN (3.7, 0.64), r$^2$SCAN~\cite{Furness2020} (3.8, 0.57), and 18 of 19 conventional meta-GGAs on both objectives. The gain is largest outside the training distribution: on molecules with bonded atom pairs absent from training, EGD lowers the NN-mGGA error from 9.1 to 2.5~\kcal\ (SM Sec.~S15).

% =====================================================================
%  OUTLOOK AND CONCLUSION
% =====================================================================
\textit{Conclusion.}---We have introduced a low-cost workflow that fits the density an XC functional would reach self-consistently, predicted in closed form from its Brillouin residual on inverse-KS orbitals of accurate reference densities, together with a stationarity loss on the predicted orbital rotation; no SCF enters the fit. For the linear MN15 form, it keeps fitting convex and yields a hybrid that improves on MN15 for held-out energies and unfitted electric-field properties and gives the most accurate densities of eight hybrids on 228 unseen density records. For a neural meta-GGA trained on six energies and 11 densities, it reverses the ``better energies, worse densities'' trend of energy-only training and surpasses SCAN in both. Because the reference density can come from any accurate method, each density system costs one Fock build, and the convex problem does not grow with the data, density-informed fitting is no longer confined to a few molecules: it now operates at the scale of the energy databases used to build modern functionals, and it can extend to systems that are hard for differentiable SCF, such as transition-metal compounds and solids. More generally, every learned model used inside a self-consistent or variational solver---Kohn-Sham and classical density functionals~\cite{Sammuller2023}, mean-field closures, and machine-learned effective Hamiltonians~\cite{Li2022DeepH}---faces the same mismatch between the state in which it is fitted and the state in which it is used. Stationarity at the reference solution, in place of differentiation through the fixed point, offers a general and inexpensive constraint for this class of problems.

\begin{acknowledgments}
X.H. was supported by the National Natural Science Foundation of China (Grant Nos. 22633005, 92477103 and 22273023), Shanghai Municipal Science and Technology Commission with Grant No. 25511102400, Shanghai Municipal Commission of Economy and Informatization with Grant No. 2026-GZL-RGZN-01015, Advanced Materials-National Science and Technology Major Project (Grant No. 2026ZD0623702), the Shanghai Frontiers Science Center of Molecule Intelligent Syntheses, and the Fundamental Research Funds for the Central Universities. We also acknowledge the Supercomputer Center of East China Normal University (ECNU Multifunctional Platform for Innovation 001) for providing computer resources.
\end{acknowledgments}

\bibliography{refs}

\end{document}

\maketitle

This supplement derives the reference-density construction, Wu--Yang
(WY) inversion, and energy (E), stationarity (G), and density-response
(D) losses implemented in this repository. The derivation applies to the
six released NN-GGA and NN-mGGA checkpoints. It gives the mathematical
definitions and the array operations needed to reproduce their training.
Section S13 gives the corresponding derivation for the linear hybrid
CDF26D, whose density constraint uses the same response construction
with exact exchange included; Section S14 defines the data sets and
evaluation metrics used in the main text; and Section S15 reports
additional results for the neural functionals. References {[}1--8{]}
identify the underlying coupled-cluster, inversion, functional, and
software methods.

All quantities are in atomic units unless stated otherwise. The
implementation uses real molecular orbitals, an aug-cc-pVTZ orbital
basis, PySCF grid level 6, and float64 arithmetic. The reference is an
all-electron UHF/UCCSD Lambda one-particle reduced density matrix
(1-RDM). Coupled-cluster amplitude response is included through the
Lambda equations; additional orbital response is absent. The training
response is a fixed-reference, diagonal-gap approximation, and the
distinction from fully coupled orbital response is derived below.

\hypertarget{s1.-notation-and-the-fixed-input-construction}{%
\subsection{S1. Notation and the fixed-input
construction}\label{s1.-notation-and-the-fixed-input-construction}}

Let \(\chi_\mu(\mathbf r)\) denote real atomic-orbital (AO) basis
functions, with overlap \(S_{\mu\nu}=\int\chi_\mu\chi_\nu\,d\mathbf r\).
Molecular-orbital (MO) coefficients satisfy \(C_\sigma^{T}SC_\sigma=I\),
where \(\sigma\in\{\uparrow,\downarrow\}\). Indices \(i,j,k\) denote
occupied spin orbitals, \(a,b,c\) virtual spin orbitals, and \(p,q\)
unrestricted MO indices. Unless a spin label is displayed, the
coupled-cluster sums run over both spins. Within each determinant spin
block, occupied orbitals have occupation one.

For an AO spin density matrix \(D_\sigma\), the density and particle
number are

\[
\rho_\sigma(\mathbf r)=\sum_{\mu\nu}
\chi_\mu(\mathbf r)D_{\sigma,\mu\nu}\chi_\nu(\mathbf r),
\qquad
N_\sigma=\operatorname{Tr}(SD_\sigma).
\tag{S1}
\]

The correlated reference density is denoted
\(\rho^{\mathrm{CC}}=\rho_\uparrow^{\mathrm{CC}}+\rho_\downarrow^{\mathrm{CC}}\).
A superscript \(0\) denotes the determinant obtained by WY inversion,
with occupied coefficients \(C^0_{o\sigma}\), virtual coefficients
\(C^0_{v\sigma}\), eigenvalues \(\epsilon^0_{p\sigma}\), and

\[
D^0_\sigma=C^0_{o\sigma}C^{0T}_{o\sigma},
\qquad D^0_\sigma S D^0_\sigma=D^0_\sigma,
\qquad
\rho^0=\rho^0_\uparrow+\rho^0_\downarrow.
\tag{S2}
\]

The correlated 1-RDM is generally not idempotent, and these determinant
matrices need not equal the correlated matrices. WY inversion targets
the spatial density or its represented moments, rather than the full
correlated 1-RDM. The residual spatial-density difference is retained in
the density loss. The reference orbitals, eigenvalues, quadrature points
and weights, and correlated targets are fixed throughout a training run.
The learned parameters are
\(\boldsymbol\theta=(\alpha,\beta,\boldsymbol\omega)\), where
\(\boldsymbol\omega\) contains the neural-network weights and biases.

The construction proceeds from UHF/UCCSD Lambda densities to WY
orbitals, then to fixed orbital/grid arrays, and finally to parameter
optimization. The CCSD energy stored with a density is
reference-preparation metadata. Energy supervision uses the reaction
targets in the selected input manifest; it is not obtained by silently
replacing those targets with CCSD total energies.

\hypertarget{s2.-ccsd-lambda-density}{%
\subsection{S2. CCSD Lambda density}\label{s2.-ccsd-lambda-density}}

The workflow requires only an accurate reference density that can be
inverted to a Kohn--Sham determinant (Section S3); the correlated method
that produces it is not part of the method. This section documents the
UCCSD Lambda densities used as references in this work.

\hypertarget{s2.1.-right-hand-amplitudes-and-the-coupled-cluster-lagrangian}{%
\subsubsection{S2.1. Right-hand amplitudes and the coupled-cluster
Lagrangian}\label{s2.1.-right-hand-amplitudes-and-the-coupled-cluster-lagrangian}}

The reference determinant is obtained from the unrestricted
Hartree--Fock equations

\[
F_\sigma^{\mathrm{HF}}
=h+J[D_\uparrow^{\mathrm{HF}}+D_\downarrow^{\mathrm{HF}}]
-K[D_\sigma^{\mathrm{HF}}],
\qquad
F_\sigma^{\mathrm{HF}}C_\sigma^{\mathrm{HF}}
=SC_\sigma^{\mathrm{HF}}\epsilon_\sigma^{\mathrm{HF}}.
\]

In a real AO basis,
\(J_{\mu\nu}[D]=\sum_{\lambda\kappa}(\mu\nu|\lambda\kappa)D_{\lambda\kappa}\)
and
\(K_{\mu\nu}[D_\sigma]=\sum_{\lambda\kappa}(\mu\lambda|\nu\kappa)D_{\sigma,\lambda\kappa}\).
The occupied up- and down-spin orbitals form the determinant used to
define excitations. The HF orbital coefficients are held fixed when
evaluating the Lambda density.

For the UHF reference determinant \(|\Phi_0\rangle\), the CCSD right
state is

\[
|\Psi_R\rangle=e^{\hat T}|\Phi_0\rangle,\qquad
\hat T=\hat T_1+\hat T_2,\qquad
\hat T_1=\sum_{ia}t_i^a a_a^\dagger a_i,\qquad
\hat T_2=\frac14\sum_{ijab}t_{ij}^{ab}
a_a^\dagger a_b^\dagger a_j a_i.
\tag{S3}
\]

The doubles amplitudes are antisymmetric under exchange of either
occupied or virtual indices. The electronic Hamiltonian and its
similarity transform are

\[
\hat H=\sum_{pq}h_{pq}a_p^\dagger a_q
+\frac14\sum_{pqrs}\langle pq\Vert rs\rangle
a_p^\dagger a_q^\dagger a_s a_r,\qquad
\overline H=e^{-\hat T}\hat H e^{\hat T}.
\tag{S4}
\]

The nuclear repulsion is an additive constant. For a Hamiltonian with at
most two-body terms, the connected Baker--Campbell--Hausdorff expansion
terminates after the fourth nested commutator:

\[
\overline H=\hat H+[\hat H,\hat T]
+\frac12[[\hat H,\hat T],\hat T]
+\frac1{6}[[[\hat H,\hat T],\hat T],\hat T]
+\frac1{24}[[[[\hat H,\hat T],\hat T],\hat T],\hat T].
\tag{S5}
\]

Projection onto the reference, singles, and doubles gives

\[
E_{\mathrm{CC}}=\langle\Phi_0|\overline H|\Phi_0\rangle,
\qquad
R_\mu(\mathbf t)=\langle\Phi_\mu|\overline H|\Phi_0\rangle=0.
\tag{S6}
\]

Here \(\mu\) labels independent singly and doubly excited determinants.
For completeness, normal ordering with respect to \(|\Phi_0\rangle\)
gives the explicit spin-orbital CCSD energy

\[
E_{\mathrm{CC}}=E_{\mathrm{HF}}
+\sum_{ia}f_{ia}t_i^a
+\frac14\sum_{ijab}\langle ij\Vert ab\rangle t_{ij}^{ab}
+\frac12\sum_{ijab}\langle ij\Vert ab\rangle t_i^a t_j^b,
\]

where \(f\) is the reference HF Fock matrix. In canonical converged HF
orbitals its occupied--virtual block vanishes, but the expression is
also valid before that simplification. The projected equations in Eq.
(S6), together with the terminating expansion in Eq. (S5), define the
singles and doubles amplitude residuals.

Because the CC energy is not variational with respect to the right-hand
amplitudes, its derivative cannot be obtained by treating the converged
amplitudes as constants in \(E_{\mathrm{CC}}\) alone. Introduce the
de-excitation operator

\[
\hat\Lambda=\hat\Lambda_1+\hat\Lambda_2,\qquad
\hat\Lambda_1=\sum_{ia}\lambda_i^a a_i^\dagger a_a,\qquad
\hat\Lambda_2=\frac14\sum_{ijab}\lambda_{ij}^{ab}
a_i^\dagger a_j^\dagger a_b a_a,
\tag{S7}
\]

and the stationary Lagrangian {[}1{]}

\[
\mathcal L_{\mathrm{CC}}(\mathbf t,\boldsymbol\lambda)
=E_{\mathrm{CC}}+\sum_\mu\lambda_\mu R_\mu
=\langle\Phi_0|(1+\hat\Lambda)\overline H|\Phi_0\rangle.
\tag{S8}
\]

Stationarity with respect to \(\boldsymbol\lambda\) recovers Eq. (S6).
Stationarity with respect to the amplitudes yields the Lambda equations.
If \(A_{\nu\mu}=\partial R_\nu/\partial t_\mu\) and
\(\eta_\mu=\partial E_{\mathrm{CC}}/\partial t_\mu\), then

\[
\sum_\nu A_{\nu\mu}\lambda_\nu=-\eta_\mu,
\quad\text{or}\quad A^T\boldsymbol\lambda=-\boldsymbol\eta,
\qquad
\langle\Phi_0|(1+\hat\Lambda)
[\overline H,\hat\tau_\mu]|\Phi_0\rangle=0.
\tag{S9}
\]

The last expression uses mutually commuting pure excitation generators
\(\hat\tau_\mu\). The left state is
\(\langle\Psi_L|=\langle\Phi_0|(1+\hat\Lambda)e^{-\hat T}\), so
\(\langle\Psi_L|\Psi_R\rangle=1\). It is not constructed by assuming
\(\Lambda=T^\dagger\).

\hypertarget{s2.2.-density-as-a-fixed-orbital-one-body-derivative}{%
\subsubsection{S2.2. Density as a fixed-orbital one-body
derivative}\label{s2.2.-density-as-a-fixed-orbital-one-body-derivative}}

Perturb the Hamiltonian by \(\eta\hat U\), where
\(\hat U=\sum_{pq}u_{pq}a_p^\dagger a_q\), and hold the orbital basis
fixed. At the stationary solution,

\[
\frac{dE_{\mathrm{CC}}}{d\eta}
=\frac{d\mathcal L_{\mathrm{CC}}}{d\eta}
=\left.\frac{\partial\mathcal L_{\mathrm{CC}}}{\partial\eta}
\right|_{\mathbf t,\boldsymbol\lambda}
=\langle\Phi_0|(1+\hat\Lambda)e^{-\hat T}
\hat U e^{\hat T}|\Phi_0\rangle.
\tag{S10}
\]

The implicit derivatives of \(\mathbf t\) and \(\boldsymbol\lambda\)
disappear from this expression because their Lagrangian derivatives
vanish. Define the unsymmetrized response density in the convention used
by PySCF,

\[
\widetilde\gamma_{pq}
=\langle\Phi_0|(1+\hat\Lambda)e^{-\hat T}
a_q^\dagger a_p e^{\hat T}|\Phi_0\rangle,
\qquad
\frac{dE_{\mathrm{CC}}}{d\eta}
=\sum_{pq}u_{pq}\widetilde\gamma_{qp}.
\tag{S11}
\]

Thus the contraction convention is
\(\operatorname{Tr}(u\widetilde\gamma)\), not an elementwise contraction
with unreversed indices. An equivalent derivation differentiates the CC
Lagrangian with respect to the one-electron integrals. For the one-body
operator \(\hat O_{pq}=a_q^\dagger a_p\), its similarity transform with
a pure excitation operator terminates as

\[
e^{-\hat T}\hat O_{pq}e^{\hat T}
=\hat O_{pq}+[\hat O_{pq},\hat T]
+\frac12[[\hat O_{pq},\hat T],\hat T].
\]

Contracting these terms with \(\langle\Phi_0|(1+\Lambda)\) and
\(|\Phi_0\rangle\), using the occupied contraction
\(\langle\Phi_0|a_i^\dagger a_j|\Phi_0\rangle=\delta_{ij}\) and the
virtual contraction
\(\langle\Phi_0|a_a a_b^\dagger|\Phi_0\rangle=\delta_{ab}\), produces
the four blocks below. In particular, occupied and virtual number
corrections pair a right excitation with a Lambda de-excitation; the
off-diagonal block also retains the double-commutator terms.

Equation (S10) includes the response of the CC amplitudes at fixed
orbitals. If the underlying HF orbitals are also allowed to vary with
the perturbation, an additional orbital-response contribution must be
included. That contribution, often obtained with a separate
orbital/Z-vector equation, is absent from the released density. The term
``Lambda density'' here therefore does not mean an orbital-relaxed
density.

\hypertarget{s2.3.-explicit-spin-orbital-ccsd-1-rdm-contractions}{%
\subsubsection{S2.3. Explicit spin-orbital CCSD 1-RDM
contractions}\label{s2.3.-explicit-spin-orbital-ccsd-1-rdm-contractions}}

For real amplitudes, define the occupied and virtual intermediates

\[
A_{ij}=\frac12\sum_{kab}\lambda_{ik}^{ab}t_{jk}^{ab},
\qquad
B_{ab}=\frac12\sum_{ijc}t_{ij}^{ca}\lambda_{ij}^{cb},
\qquad
Z_{ae}=B_{ae}+\sum_m t_m^a\lambda_m^e.
\tag{S12}
\]

Using the operator ordering in Eqs. (S3) and (S7), the four correlation
blocks are

\[
d^{oo}_{ij}=-\sum_a\lambda_i^a t_j^a-A_{ij},
\qquad
d^{vv}_{ab}=\sum_i t_i^a\lambda_i^b+B_{ab},
\qquad
d^{ov}_{ia}=\lambda_i^a,
\tag{S13}
\]

\[
d^{vo}_{ai}
=t_i^a+\sum_{me}t_{im}^{ae}\lambda_m^e
-\sum_m A_{mi}t_m^a-\sum_e t_i^e Z_{ae}.
\tag{S14}
\]

The complete unsymmetrized matrix, including the reference determinant,
is

\[
\widetilde\gamma=
\begin{pmatrix}
I_o+d^{oo}&d^{ov}\\
d^{vo}&d^{vv}
\end{pmatrix}.
\tag{S15}
\]

The identity in the occupied block is essential: the four \(d\) blocks
alone contain only the correction and off-diagonal terms. The factors of
\(1/2\) in Eq. (S12) follow from antisymmetric doubles amplitudes and
the two equivalent contractions of a spectator occupied or virtual
index. Equation (S14) contains the linear singles term, the
doubles--Lambda-singles term, and the two intermediate contractions;
omitting the last two terms would not give the CCSD response density.

For unrestricted CCSD, the nonzero tensors are
\(t_1^\uparrow,t_1^\downarrow\) and
\(t_2^{\uparrow\uparrow},t_2^{\uparrow\downarrow},t_2^{\downarrow\downarrow}\),
with corresponding Lambda tensors. Equations (S12)--(S15) already
include all spin cases when interpreted as spin-orbital sums. For
example, using lowercase indices for up-spin orbitals and uppercase
indices for down-spin orbitals,

\[
A^\uparrow_{ij}
=\frac12\sum_{kab}\lambda^{\uparrow\uparrow}_{ik}{}^{ab}
t^{\uparrow\uparrow}_{jk}{}^{ab}
+\sum_{KaB}\lambda^{\uparrow\downarrow}_{iK}{}^{aB}
t^{\uparrow\downarrow}_{jK}{}^{aB},
\tag{S16}
\]

\[
B^\uparrow_{ab}
=\frac12\sum_{ijc}t^{\uparrow\uparrow}_{ij}{}^{ac}
\lambda^{\uparrow\uparrow}_{ij}{}^{bc}
+\sum_{iJB}t^{\uparrow\downarrow}_{iJ}{}^{aB}
\lambda^{\uparrow\downarrow}_{iJ}{}^{bB}.
\tag{S17}
\]

The opposite-spin terms have coefficient one. They arise from the two
spin arrangements represented by the antisymmetric spin-orbital sum;
applying another factor of \(1/2\) to them would be incorrect.
Interchanging the spin labels gives the down-spin blocks. The
doubles--Lambda-singles term in \(d^{vo,\uparrow}_{ai}\) similarly
contains both
\(\sum_{me}t^{\uparrow\uparrow}_{im}{}^{ae}\lambda^\uparrow_m{}^e\) and
\(\sum_{ME}t^{\uparrow\downarrow}_{iM}{}^{aE}\lambda^\downarrow_M{}^E\).
The remaining contractions follow Eqs. (S13) and (S14) within each spin
block. This is the decomposition used in PySCF's UCCSD RDM
implementation {[}8{]}.

For real orbitals and the symmetric one-electron perturbations relevant
to a scalar density, only the symmetric part contributes. PySCF returns

\[
\gamma=\frac12(\widetilde\gamma+\widetilde\gamma^T),
\qquad
\gamma_{oo}=I_o+\frac12(d^{oo}+d^{ooT}),
\qquad
\gamma_{ov}=\frac12(d^{ov}+d^{voT}),
\tag{S18}
\]

with \(\gamma_{vo}=\gamma_{ov}^T\) and
\(\gamma_{vv}=(d^{vv}+d^{vvT})/2\). Hermitization is not an
orbital-response correction. The antisymmetric part of the real
unsymmetrized matrix makes no contribution to \(\rho(\mathbf r)\)
because the orbital products are symmetric. Neither Hermitization nor
particle-number conservation is, by itself, a proof of positive
semidefiniteness for a truncated coupled-cluster 1-RDM.

\hypertarget{s2.4.-ao-transformation-quadrature-and-implementation}{%
\subsubsection{S2.4. AO transformation, quadrature, and
implementation}\label{s2.4.-ao-transformation-quadrature-and-implementation}}

The separate spin matrices are transformed with the UHF MO coefficients:

\[
D_\sigma^{\mathrm{CC}}=C_\sigma^{\mathrm{HF}}
\gamma_\sigma C_\sigma^{\mathrm{HF}T},
\qquad
\rho^{\mathrm{CC}}(\mathbf r)
=\sum_\sigma\sum_{\mu\nu}
D_{\sigma,\mu\nu}^{\mathrm{CC}}\chi_\mu(\mathbf r)\chi_\nu(\mathbf r).
\tag{S19}
\]

The particle-number identity follows from MO orthonormality:

\[
\operatorname{Tr}(SD_\sigma^{\mathrm{CC}})
=\operatorname{Tr}(\gamma_\sigma)=N_\sigma.
\tag{S20}
\]

The trace cancellation can also be checked directly in Eqs. (S12) and
(S13): the occupied and virtual singles contractions cancel, as do the
corresponding doubles contractions. On a finite grid the integral
\(\sum_\ell w_\ell\rho^{\mathrm{CC}}(\mathbf r_\ell)\) approximates this
AO trace and has a separate quadrature error.

\nolinkurl{scripts/prepare_references.py} runs UHF, all-electron UCCSD
with \texttt{frozen=0}, and \nolinkurl{solve_lambda}, followed by
\nolinkurl{make_rdm1(ao_repr=True)} in PySCF {[}7,8{]}. The output
\nolinkurl{dm_lambda} has shape \((2,N_{\mathrm{AO}},N_{\mathrm{AO}})\).
No triples contribution, density fitting, or additional orbital-response
density is added. For a one-electron system the density is the UHF
one-electron limit. The reference script requires converged
UHF/CCSD/Lambda calculations and a finite AO density with
spin-electron-count error below \(10^{-7}\). Its UHF energy tolerance is
\(10^{-12}\,E_h\); CCSD uses an energy tolerance of \(10^{-10}\,E_h\),
an amplitude-norm tolerance of \(10^{-8}\), and up to 150 iterations.
The explicit CCSD 1-RDM contractions above can be evaluated from stored
amplitudes without re-solving a production calculation.

\hypertarget{s3.-regularized-finite-basis-wuyang-inversion}{%
\subsection{S3. Regularized finite-basis Wu--Yang
inversion}\label{s3.-regularized-finite-basis-wuyang-inversion}}

\hypertarget{s3.1.-variational-dual-and-the-local-potential}{%
\subsubsection{S3.1. Variational dual and the local
potential}\label{s3.1.-variational-dual-and-the-local-potential}}

For a fixed occupation pattern, the noninteracting ground-state energy
in a local potential is

\[
E_s[v]=\min_{\Phi}\left\{
T_s[\Phi]+\int v(\mathbf r)\rho_\Phi(\mathbf r)\,d\mathbf r
\right\}.
\tag{S21}
\]

It is concave in \(v\) because it is the minimum of functions that are
affine in the potential. The density-to-potential dual is

\[
T_s^{\mathrm{dual}}[\rho^{\mathrm{CC}}]
=\sup_v\left\{
E_s[v]-\int v(\mathbf r)\rho^{\mathrm{CC}}(\mathbf r)\,d\mathbf r
\right\}.
\]

Where the supremum is attained on a differentiable branch, its potential
derivative is the difference between the determinant and target
densities {[}2{]}. In the finite orbital and potential bases used here,
the optimized potential has the form

\[
v_\sigma(\mathbf r)
=v_{\mathrm{ext}}(\mathbf r)
+v_{\mathrm{FA}}\rho^{\mathrm{CC}} (`\mathbf r`)
+\sum_t b_{\sigma t}\eta_t(\mathbf r),
\qquad
v_{\mathrm{FA}}[\rho^{\mathrm{CC}}]
=\frac{N-1}{N}v_H[\rho^{\mathrm{CC}}].
\tag{S22}
\]

The Fermi--Amaldi guide and the correlated density are held fixed while
varying \(\mathbf b_\sigma\). The potential basis functions \(\eta_t\)
use aug-cc-pVTZ in the released preparation routine. The inverse
potential is a local one-electron potential; it is not assumed to be the
potential of the subsequently learned functional.

Define the potential matrices and target moments

\[
(V_t)_{\mu\nu}
=\int\chi_\mu\chi_\nu\eta_t\,d\mathbf r,\qquad
n_{\sigma t}^{\mathrm{tar}}
=\operatorname{Tr}(V_tD_\sigma^{\mathrm{CC}}),
\qquad
h_\sigma(\mathbf b_\sigma)=h_{\mathrm{guide}}+\sum_t b_{\sigma t}V_t.
\tag{S23}
\]

The generalized eigenproblem is
\(h_\sigma C_\sigma=SC_\sigma\epsilon_\sigma\). The code obtains \(V_t\)
from three-center overlap integrals, not from a density-fitting
approximation to the two-electron Coulomb operator.

\hypertarget{s3.2.-spatial-smoothing-and-occupation-factors}{%
\subsubsection{S3.2. Spatial smoothing and occupation
factors}\label{s3.2.-spatial-smoothing-and-occupation-factors}}

For one optimized channel \(c\), let \(f_c\) be its orbital-occupation
factor and \(D_c^{\mathrm{tar}}\) its target. For an unrestricted spin
channel, \(f_c=1\) and \(D_c^{\mathrm{tar}}=D_\sigma^{\mathrm{CC}}\).
For a restricted closed-shell calculation, \(f_c=2\) and
\(D_c^{\mathrm{tar}}=D_\uparrow^{\mathrm{CC}}+D_\downarrow^{\mathrm{CC}}\).
Dropping guide-potential constants that do not depend on
\(\mathbf b_c\), minimization of the negative regularized dual gives

\[
J_c(\mathbf b_c)
=-f_c\sum_{i=1}^{n_{oc}}\epsilon_{ic}(\mathbf b_c)
+\sum_t b_{ct}n_{ct}^{\mathrm{tar}}
+\lambda_{\mathrm{WY}}\mathbf b_c^T R\mathbf b_c,
\qquad \lambda_{\mathrm{WY}}=10^{-5},
\tag{S24}
\]

where

\[
R_{tu}=\int\nabla\eta_t\cdot\nabla\eta_u\,d\mathbf r
=2\langle\eta_t|-\tfrac12\nabla^2|\eta_u\rangle,
\qquad
\mathbf b_c^TR\mathbf b_c
=\int\left|\nabla\sum_t b_{ct}\eta_t\right|^2d\mathbf r.
\tag{S25}
\]

The spatial penalty suppresses rapidly varying potential components in a
finite basis {[}3{]}. It is distinct from neural-network weight decay,
which is zero in these fits. In the restricted implementation there is
one optimized channel and one penalty in Eq. (S24); its resulting
spatial orbitals are then copied into two spin blocks. The penalty is
not silently multiplied by two. A spin-summed target used in this
restricted calculation does not separately fit any difference between
the two UHF spin densities.

An additive constant in the potential shifts the occupied eigenvalue sum
and the target-density integral by the same particle-number factor,
leaving the unregularized dual unchanged. A gradient penalty also leaves
an exact constant unchanged. The present finite Gaussian potential basis
does not include an exact spatial constant; the implemented Cholesky
transformation uses its full positive potential-gradient metric. This
numerical parameterization should not be interpreted as a general proof
of uniqueness of a continuum inverse potential.

\hypertarget{s3.3.-gradient-density-moments-and-hessian}{%
\subsubsection{S3.3. Gradient, density moments, and
Hessian}\label{s3.3.-gradient-density-moments-and-hessian}}

For an isolated eigenvalue and fixed occupations, the Hellmann--Feynman
identity gives \(\partial\epsilon_i/\partial b_t=C_i^TV_tC_i\).
Therefore

\[
\frac{\partial J_c}{\partial b_t}
=n_{ct}^{\mathrm{tar}}
-f_c\sum_i C_{ic}^TV_tC_{ic}
+2\lambda_{\mathrm{WY}}(R\mathbf b_c)_t.
\tag{S26}
\]

The first two terms are the target-minus-determinant density moments in
the potential basis. At a regularized optimum those moments differ by
\(-2\lambda_{\mathrm{WY}}R\mathbf b_c\); a converged optimizer therefore
does not imply exact pointwise density reproduction. Even without
regularization, only the represented potential-basis moments are
constrained. Correlated densities also need not be exactly representable
by a pure determinant in the selected finite orbital basis.

For positive occupied--virtual gaps, first-order eigenvector
perturbation gives

\[
\frac{\partial C_{ic}}{\partial b_u}
=\sum_{a\in\mathrm{vir}}
C_{ac}\frac{C_{ac}^TV_uC_{ic}}{\epsilon_{ic}-\epsilon_{ac}}
+\text{occupied-space gauge terms}.
\tag{S27}
\]

The occupied-space terms cancel when differentiating the occupied
projector. Differentiating Eq. (S26) then yields

\[
\frac{\partial^2J_c}{\partial b_t\partial b_u}
=2f_c\sum_{ia}
\frac{(C_{ac}^TV_tC_{ic})(C_{ac}^TV_uC_{ic})}
{\epsilon_{ac}-\epsilon_{ic}}
+2\lambda_{\mathrm{WY}}R_{tu}.
\tag{S28}
\]

The factor of two comes from differentiating the bra and ket in each
density moment. With positive gaps and a positive potential metric, this
is a positive semidefinite response term plus a positive regularizer.
Equation (S28) is a derivative on a branch with a fixed occupied
subspace. At an occupied--virtual degeneracy the usual
differentiable-branch argument breaks down. The implementation floors
these inverse-solver denominators at \(10^{-10}\,E_h\) as a numerical
safeguard; this is separate from the \(0.05\,E_h\) training-response
floor introduced in Section S6.

\hypertarget{s3.4.-ao-orthogonalization-and-potential-metric-coordinates}{%
\subsubsection{S3.4. AO orthogonalization and potential-metric
coordinates}\label{s3.4.-ao-orthogonalization-and-potential-metric-coordinates}}

The solver uses full-rank Cholesky transformations. With \(S=L_SL_S^T\),
let \(X=L_S^{-T}\) so that \(X^TSX=I\). With \(R=L_RL_R^T\), define

\[
Q=L_R^{-T},\qquad \mathbf b=Q\mathbf z,\qquad
\mathbf z=L_R^T\mathbf b,\qquad
\widetilde h=X^Th_{\mathrm{guide}}X,\qquad
\widetilde V_t=\sum_u Q_{ut}X^TV_uX,\qquad
\widetilde{\mathbf n}=Q^T\mathbf n^{\mathrm{tar}}.
\tag{S29}
\]

Then \(\mathbf b^TR\mathbf b=\mathbf z^T\mathbf z\), and the same
objective becomes

\[
J_c(\mathbf z)
=-f_c\sum_i\epsilon_{ic}(\mathbf z)
+\mathbf z^T\widetilde{\mathbf n}
+\lambda_{\mathrm{WY}}\mathbf z^T\mathbf z,
\qquad
\nabla_{\mathbf b}J_c=L_R\nabla_{\mathbf z}J_c.
\tag{S30}
\]

The transformed gradient and Hessian follow Eqs. (S26) and (S28) with
the transformed potential matrices and \(R\rightarrow I\). The Newton
step is formed in \(\mathbf z\) coordinates; Hessian eigenvalues are
floored at \(2\lambda_{\mathrm{WY}}\), and a backtracking line search
controls the step. BFGS followed by Newton polishing is attempted when
the raw gradient remains too large. The transformations change
conditioning without deleting AO or potential directions.

\nolinkurl{src/ccxc/wy_polish.py} reports convergence from the gradient in
the original \(\mathbf b\) coordinates, not just the whitened-coordinate
norm or an optimizer success flag. Its final checks require raw gradient
norm below \(10^{-5}\), maximum local eigenproblem residual below
\(10^{-7}\), maximum MO orthogonality error below \(10^{-7}\), and
spin-electron-count error below \(10^{-7}\), with finite arrays. The
inner Newton loop targets a raw gradient norm below \(10^{-8}\). These
are numerical inversion conditions; they do not certify exact
representability of the correlated density.

The returned determinant is Eq. (S2), using the lowest occupied
eigenvectors of the optimized local Hamiltonian. Its coefficients and
all occupied/virtual eigenvalues are retained. The eigenvalues provide
the fixed response denominators in training. The meta-GGA kinetic
density is constructed from these occupied determinant orbitals, rather
than from a correlated kinetic-density field inferred from the CCSD
1-RDM.

\hypertarget{s4.-neural-functional-and-its-generalized-kohnsham-operator}{%
\subsection{S4. Neural functional and its generalized Kohn--Sham
operator}\label{s4.-neural-functional-and-its-generalized-kohnsham-operator}}

\hypertarget{s4.1.-energy-raw-invariants-and-descriptors}{%
\subsubsection{S4.1. Energy, raw invariants, and
descriptors}\label{s4.1.-energy-raw-invariants-and-descriptors}}

For a determinant density \(D=D_\uparrow+D_\downarrow\), the candidate
total energy is

\[
E_{\boldsymbol\theta}[D_\uparrow,D_\downarrow]
=\operatorname{Tr}(hD)+\frac12\operatorname{Tr}(J[D]D)
+E_{\mathrm{nuc}}+E_{xc,\boldsymbol\theta},
\qquad
E_{xc,\boldsymbol\theta}
=\alpha E_{xc}^{\mathrm{base}}
+\int q_{\boldsymbol\theta}(\mathbf x(\mathbf r))\,d\mathbf r,
\tag{S31}
\]

where \(h\) contains the physical kinetic and nuclear-attraction
operators and \(J[D]\) is the Hartree matrix of the determinant density.
The Fermi--Amaldi inversion guide does not replace this Hartree
contribution. The base is PBE {[}4{]} for NN-GGA and SCAN {[}5{]} for
NN-mGGA. The local neural correction is

\[
q_{\boldsymbol\theta}(\mathbf x)
=\beta n f_{\boldsymbol\omega}(\mathbf d(\mathbf x)),\qquad
\mathbf x=(n_\uparrow,n_\downarrow,
\sigma_{\uparrow\uparrow},\sigma_{\uparrow\downarrow},
\sigma_{\downarrow\downarrow},\tau_\uparrow,\tau_\downarrow),
\qquad
\sigma_{\sigma\sigma'}=\nabla n_\sigma\cdot\nabla n_{\sigma'}.
\tag{S32}
\]

For GGA the final two entries are omitted. For the determinant,

\[
n_\sigma=\sum_{i\in o_\sigma}\phi_{i\sigma}^2,\qquad
\nabla n_\sigma=2\sum_i\phi_{i\sigma}\nabla\phi_{i\sigma},
\qquad
\tau_\sigma=\frac12\sum_i|\nabla\phi_{i\sigma}|^2.
\tag{S33}
\]

The regularized descriptors used in the code are specified completely by

\[
\begin{aligned}
n&=n_\uparrow+n_\downarrow,\qquad
\bar n=\max(n,10^{-18}),\qquad
\Sigma=\sigma_{\uparrow\uparrow}
+2\sigma_{\uparrow\downarrow}+\sigma_{\downarrow\downarrow},\\
s&=\frac{\sqrt{\max(\Sigma,10^{-36})}}
{2(3\pi^2)^{1/3}\bar n^{4/3}},\qquad
\zeta=\frac{n_\uparrow-n_\downarrow}{n+10^{-18}},\\
\tau_W&=\frac{\Sigma}{8\bar n},\qquad
\tau_{\mathrm{unif}}
=\frac3{10}(6\pi^2)^{2/3}
\left[\max(n_\uparrow,10^{-18})^{5/3}
+\max(n_\downarrow,10^{-18})^{5/3}\right],\\
a_\tau&=\frac{\max(\tau_\uparrow+\tau_\downarrow-\tau_W,0)}
{\tau_{\mathrm{unif}}},\\
\mathbf d_{\mathrm{GGA}}
&=\bigl(\ln[1+\max(n,0)],\zeta,\ln(1+s)\bigr),\\
\mathbf d_{\mathrm{mGGA}}
&=\bigl(\ln[1+\max(n,0)],\zeta,\ln(1+s),\ln(1+a_\tau)\bigr).
\end{aligned}
\tag{S34}
\]

The numerical constant \(2(3\pi^2)^{1/3}\) is 6.187335452560271 in the
implementation. Logarithmic density descriptors use the numerical values
in atomic units. The signed spin descriptor is retained without an
additional spin-exchange symmetrization. The spin-dependent uniform-gas
convention makes \(a_\tau=1\) for a uniform gas, including a
spin-polarized gas, away from the numerical floors. The Pauli-density
clamp and density floors are piecewise extensions; global smoothness at
their boundaries is not asserted.

The network has three hidden layers of width 32, each followed by
\(\operatorname{Softplus}(z)=\ln(1+e^z)\), and a scalar linear output
without bias. The input dimension is three or four. Thus

\[
N_{\mathrm{par}}
=(d_{\mathrm{in}}+1)32+2(32+1)32+32+2
=
\begin{cases}
2274,&d_{\mathrm{in}}=3,\\
2306,&d_{\mathrm{in}}=4.
\end{cases}
\tag{S35}
\]

The final two parameters are \(\alpha\) and \(\beta\). The GGA
architecture follows the descriptor/network construction of Kasim and
Vinko {[}6{]}. The SCAN extension adds the fourth descriptor and all
corresponding kinetic-density derivatives.

The neural contribution and its invariant derivatives are set to zero at
grid points with \(n\le10^{-20}\). This mask applies to the neural
correction; the base functional is evaluated by LibXC. Since the
reference invariants are fixed during training, the grid mask is also
fixed during a parameter update.

\hypertarget{s4.2.-functional-differentiation-and-the-ao-matrix}{%
\subsubsection{S4.2. Functional differentiation and the AO
matrix}\label{s4.2.-functional-differentiation-and-the-ao-matrix}}

Let
\(e(\mathbf x)=\alpha e_{xc}^{\mathrm{base}}(\mathbf x)+q_{\boldsymbol\theta}(\mathbf x)\)
denote the XC energy per volume. Define
\(e_{n_\sigma}=\partial e/\partial n_\sigma\),
\(e_{\sigma_{\sigma\sigma'}}=\partial e/\partial\sigma_{\sigma\sigma'}\),
and \(e_{\tau_\sigma}=\partial e/\partial\tau_\sigma\). Variation of the
gradient invariants gives

\[
\mathbf u_\sigma
=2e_{\sigma_{\sigma\sigma}}\nabla n_\sigma
+e_{\sigma_{\uparrow\downarrow}}\nabla n_{\bar\sigma},
\qquad
\delta E_{xc}
=\sum_\sigma\int
\left[e_{n_\sigma}\delta n_\sigma
+\mathbf u_\sigma\cdot\nabla\delta n_\sigma
+e_{\tau_\sigma}\delta\tau_\sigma\right]d\mathbf r.
\tag{S36}
\]

The mixed-spin gradient invariant appears once in \(\mathbf x\),
explaining the absence of an extra factor of two in its contribution to
\(\mathbf u_\sigma\). In an AO basis,

\[
\begin{aligned}
(V_{xc,\sigma})_{\mu\nu}
=\int\bigg[
&e_{n_\sigma}\chi_\mu\chi_\nu
+\mathbf u_\sigma\cdot
(\chi_\nu\nabla\chi_\mu+\chi_\mu\nabla\chi_\nu)\\
&+\frac12e_{\tau_\sigma}
\nabla\chi_\mu\cdot\nabla\chi_\nu
\bigg]d\mathbf r,\qquad
F_\sigma=h+J[D]+V_{xc,\sigma}.
\end{aligned}
\tag{S37}
\]

The \(1/2\) multiplying the kinetic-density term follows from the
definition of \(\tau_\sigma\) in Eq. (S33). With vanishing boundary
terms, the corresponding generalized Kohn--Sham (GKS) operator can be
written

\[
\hat F_\sigma
=-\frac12\nabla\cdot
\left[(1+e_{\tau_\sigma})\nabla\right]
+v_{\mathrm{ext}}+v_H+e_{n_\sigma}
-\nabla\cdot\mathbf u_\sigma.
\tag{S38}
\]

For a GGA, \(e_{\tau_\sigma}=0\). For a meta-GGA, neglecting this term
changes both the self-consistent operator and the training stationarity
residual. There is no density-Laplacian dependence in the released
models.

PySCF's XC callback returns the energy per particle
\(\varepsilon_{xc}=e/n\), while its potential and kernel entries
correspond to derivatives of the energy per volume with respect to the
density invariants. The code therefore differentiates
\(q_{\boldsymbol\theta}=\beta n f_{\boldsymbol\omega}\), including the
prefactor \(n\), and divides by \(n\) only when constructing the
callback energy-per-particle output. Differentiating the per-particle
neural output as if it were the energy density would give incorrect
potential terms.

\hypertarget{s4.3.-restricted-mapping-and-second-derivatives}{%
\subsubsection{S4.3. Restricted mapping and second
derivatives}\label{s4.3.-restricted-mapping-and-second-derivatives}}

For a restricted density, define \(\mathbf y=(n,\sigma,\tau)\) with
\(\sigma=|\nabla n|^2\). The spin invariants are
\(\mathbf x=T\mathbf y\), where

\[
T=
\begin{pmatrix}
1/2&0&0\\
1/2&0&0\\
0&1/4&0\\
0&1/4&0\\
0&1/4&0\\
0&0&1/2\\
0&0&1/2
\end{pmatrix},
\qquad
\nabla_{\mathbf y}q=T^T\nabla_{\mathbf x}q,\qquad
\nabla_{\mathbf y}^2q=T^T(\nabla_{\mathbf x}^2q)T.
\tag{S39}
\]

The GGA mapping is the upper five rows and first two columns. This
transformation is applied to the neural derivatives, and the
correspondingly scaled LibXC base derivatives are added. Autograd
supplies first and second invariant derivatives, which the callback
packs into the PySCF GGA/MGGA blocks. Second derivatives are available
for SCF response/Newton operations; the training loss below uses a
separate mixed invariant--parameter derivative.

\hypertarget{s5.-orbital-rotations-and-the-projected-energy-gradient}{%
\subsection{S5. Orbital rotations and the projected energy
gradient}\label{s5.-orbital-rotations-and-the-projected-energy-gradient}}

\hypertarget{s5.1.-orthogonality-preserving-variations}{%
\subsubsection{S5.1. Orthogonality-preserving
variations}\label{s5.1.-orthogonality-preserving-variations}}

For each spin, partition the reference orbitals as
\(C^0_\sigma=(C^0_{o\sigma},C^0_{v\sigma})\) and introduce a real
antisymmetric generator

\[
K_\sigma=
\begin{pmatrix}
0&-\kappa_\sigma^T\\
\kappa_\sigma&0
\end{pmatrix},
\qquad
C_\sigma(\kappa_\sigma)=C^0_\sigma e^{K_\sigma},
\qquad
\delta C_{o\sigma}=C^0_{v\sigma}\kappa_\sigma.
\tag{S40}
\]

The exponential preserves \(C_\sigma^TSC_\sigma=I\). To first order,

\[
\delta D_\sigma
=C^0_{v\sigma}\kappa_\sigma C^{0T}_{o\sigma}
+C^0_{o\sigma}\kappa_\sigma^TC^{0T}_{v\sigma},
\qquad
\delta E=\sum_\sigma\operatorname{Tr}(F_\sigma\delta D_\sigma)
=2\sum_{\sigma ai}F^\sigma_{ai}\kappa^\sigma_{ai}.
\tag{S41}
\]

It follows that the independent-spin orbital gradient is

\[
g^\sigma_{ai}
=\left.\frac{\partial E}{\partial\kappa^\sigma_{ai}}\right|_0
=2F^\sigma_{ai},
\qquad
F^\sigma_{ai}
=C^{0T}_{a\sigma}F_\sigma[D^0]C^0_{i\sigma}.
\tag{S42}
\]

Restricted determinants are represented in training by two equal spin
blocks, each with unit occupations. If a single shared spatial rotation
were used instead, its gradient would contain the sum of both spin
contributions and would be \(4F_{ai}\) for equal spin operators. That
shared-coordinate factor is not substituted into the independent-spin
vectors used in the code.

\hypertarget{s5.2.-grid-expressions-for-the-variations-and-the-tau-term}{%
\subsubsection{S5.2. Grid expressions for the variations and the tau
term}\label{s5.2.-grid-expressions-for-the-variations-and-the-tau-term}}

The density and kinetic-density variations induced by Eq. (S40) are

\[
\delta n_\sigma
=2\sum_{ai}\phi^0_{a\sigma}\phi^0_{i\sigma}\kappa^\sigma_{ai},
\qquad
\nabla\delta n_\sigma
=2\sum_{ai}
(\phi^0_{i\sigma}\nabla\phi^0_{a\sigma}
+\phi^0_{a\sigma}\nabla\phi^0_{i\sigma})\kappa^\sigma_{ai},
\tag{S43}
\]

\[
\delta\tau_\sigma
=\sum_{ai}\nabla\phi^0_{a\sigma}\cdot
\nabla\phi^0_{i\sigma}\kappa^\sigma_{ai}.
\tag{S44}
\]

There is no factor of two in Eq. (S44), because it cancels the \(1/2\)
in \(\tau_\sigma\). Substituting Eqs. (S43) and (S44) into Eq. (S36)
gives the XC part of the projected gradient directly:

\[
\begin{aligned}
g^{xc,\sigma}_{ai}
=\int\bigg[
&2e_{n_\sigma}\phi^0_{a\sigma}\phi^0_{i\sigma}\\
&+2\mathbf u_\sigma\cdot
(\phi^0_{i\sigma}\nabla\phi^0_{a\sigma}
+\phi^0_{a\sigma}\nabla\phi^0_{i\sigma})\\
&+e_{\tau_\sigma}
\nabla\phi^0_{a\sigma}\cdot\nabla\phi^0_{i\sigma}
\bigg]d\mathbf r.
\end{aligned}
\tag{S45}
\]

This is exactly \(2C_a^TV_{xc,\sigma}C_i\) from Eq. (S37). The GGA
projection in \nolinkurl{src/ccxc/xcnn_tensor_orbitals.py} forms the
first two lines. \nolinkurl{src/ccxc/mgga_projection.py} adds the final
line with the coefficient shown, rather than multiplying the tau term by
another factor of two.

\hypertarget{s5.3.-cached-fixed-orbital-energy-and-gradient}{%
\subsubsection{S5.3. Cached fixed-orbital energy and
gradient}\label{s5.3.-cached-fixed-orbital-energy-and-gradient}}

For each system, the parameter-independent energy and orbital-gradient
baseline is

\[
E_m^{(0)}
=\operatorname{Tr}(hD_m^0)
+\frac12\operatorname{Tr}(J[D_m^0]D_m^0)+E_{\mathrm{nuc},m},
\qquad
g_{m,ai\sigma}^{(0)}
=2C_{ma\sigma}^{0T}(h+J[D_m^0])C^0_{mi\sigma}.
\tag{S46}
\]

The arrays also store \(E^{\mathrm{base}}_{xc,m}\) and
\(g^{\mathrm{base}}_m\). On the quadrature grid,

\[
E_m(\boldsymbol\theta)
=E_m^{(0)}+\alpha E^{\mathrm{base}}_{xc,m}
+\sum_\ell w_{m\ell}q_{\boldsymbol\theta}(\mathbf x_{m\ell}),
\qquad
g_m(\boldsymbol\theta)
=g_m^{(0)}+\alpha g_m^{\mathrm{base}}+g_m^{\mathrm{NN}}.
\tag{S47}
\]

The PBE fields are named \texttt{epbe}/\texttt{gpbe}; the SCAN fields
are \texttt{escan}/\texttt{gscan}. Building a meta-GGA point recomputes
the SCAN baseline and adds the occupied-orbital kinetic density. Reusing
PBE baseline fields as SCAN fields would change Eq. (S47).

\hypertarget{s6.-from-the-orbital-residual-to-the-implemented-density-response}{%
\subsection{S6. From the orbital residual to the implemented density
response}\label{s6.-from-the-orbital-residual-to-the-implemented-density-response}}

\hypertarget{s6.1.-exact-local-response-and-the-approximation-used-here}{%
\subsubsection{S6.1. Exact local response and the approximation used
here}\label{s6.1.-exact-local-response-and-the-approximation-used-here}}

A local expansion of the candidate energy about the fixed determinant is

\[
E_{\boldsymbol\theta}(\boldsymbol\kappa)
=E_{\boldsymbol\theta}(0)
+g_{\boldsymbol\theta}^T\boldsymbol\kappa
+\frac12\boldsymbol\kappa^TH_{\boldsymbol\theta}\boldsymbol\kappa
+O(\|\boldsymbol\kappa\|^3).
\tag{S48}
\]

If the Hessian on the allowed orbital-rotation space is invertible, the
stationary point of the quadratic model satisfies
\(H_{\boldsymbol\theta}\boldsymbol\kappa=-g_{\boldsymbol\theta}\).
Interpreting it as a local minimum additionally requires positive
curvature and a sufficiently small displacement. The full candidate
Hessian includes changes in Hartree and XC potentials induced by the
density/orbital variation, as well as the appropriate candidate-Fock
terms. Solving that coupled problem is not part of the released training
loss.

For a noninteracting reference operator diagonal in the WY orbitals, add
an off-diagonal perturbation \(F_{ai}\) and retain the reference
diagonal curvature. Rotating one occupied orbital into a virtual orbital
then changes its one-electron energy by

\[
\delta E_s
=2F_{ai}\kappa_{ai}
+(\epsilon_a^0-\epsilon_i^0)\kappa_{ai}^2+O(\kappa_{ai}^3).
\tag{S49}
\]

Equivalently, first-order eigenvector perturbation by an off-diagonal
field gives \(\kappa_{ai}=-F_{ai}/(\epsilon_a^0-\epsilon_i^0)\). Because
\(C_a^{0T}h_{\mathrm{WY}}C_i^0=0\), the occupied--virtual element of the
perturbation \(F_{\boldsymbol\theta}[D^0]-h_{\mathrm{WY}}\) is precisely
the candidate \(F_{ai}\). The diagonal reference curvature is
\(2(\epsilon_a^0-\epsilon_i^0)\) in the independent-spin coordinates of
Eq. (S42).

The implementation retains this reference curvature, floors small gaps,
and holds the resulting preconditioner fixed:

\[
\Delta_{ai\sigma}
=\max(\epsilon^0_{a\sigma}-\epsilon^0_{i\sigma},0.05\,E_h),
\qquad
p_{ai\sigma}=\frac1{2\Delta_{ai\sigma}},
\qquad
P=\operatorname{diag}(p_{ai\sigma}),
\qquad
\boldsymbol\kappa_{\mathrm{pred}}=-Pg_{\boldsymbol\theta}.
\tag{S50}
\]

This substitutes fixed WY gaps for the full parameter-dependent
curvature. The induced Hartree/XC kernel is omitted, and the gaps are
not updated with the model. Equation (S50) is consequently a
diagonal-gap, independent-particle response surrogate, not a solution of
the full coupled-perturbed Kohn--Sham equations. Near degeneracies the
gap floor also changes the formal noninteracting response.

\hypertarget{s6.2.-linear-density-prediction-and-conservation}{%
\subsubsection{S6.2. Linear density prediction and
conservation}\label{s6.2.-linear-density-prediction-and-conservation}}

Combining Eqs. (S43) and (S50) gives the total predicted density

\[
\rho_{\mathrm{pred}}(\mathbf r)
=\rho^0(\mathbf r)
+2\sum_{\sigma ai}
\phi^0_{a\sigma}(\mathbf r)\phi^0_{i\sigma}(\mathbf r)
\kappa^{\mathrm{pred}}_{ai\sigma}
=\rho^0(\mathbf r)
-2\sum_{\sigma ai}p_{ai\sigma}g_{ai\sigma}
\phi^0_{a\sigma}(\mathbf r)\phi^0_{i\sigma}(\mathbf r).
\tag{S51}
\]

Occupied--virtual orthogonality guarantees

\[
\int(\rho_{\mathrm{pred}}-\rho^0)\,d\mathbf r
=2\sum_{\sigma ai}\kappa^{\mathrm{pred}}_{ai\sigma}
\langle\phi^0_{a\sigma}|\phi^0_{i\sigma}\rangle=0.
\tag{S52}
\]

The corresponding numerical quadrature integral may have a small
integration error. Positivity is not guaranteed because Eq. (S51)
retains only the first-order density change. The code does not clip or
renormalize the predicted density. It records the minimum predicted
density, rotation norm, maximum rotation, and integrated density change
so that the scale of the linear approximation can be inspected.

For later use, define the grid-to-rotation matrix and the residual

\[
B_{\ell,ai\sigma}
=2\phi^0_{a\sigma}(\mathbf r_\ell)\phi^0_{i\sigma}(\mathbf r_\ell),
\qquad
W=\operatorname{diag}(w_\ell),
\qquad
d_\ell=\rho^0(\mathbf r_\ell)-\rho^{\mathrm{CC}}(\mathbf r_\ell),
\qquad
r=d-BPg.
\tag{S53}
\]

The columns of \(B\) include both spin channels, so the spin density
changes are added before any density error is squared. The array \(d\)
contains the finite inversion mismatch. Setting \(d=0\) would define a
different training objective.

\hypertarget{s7.-energy-loss-reaction-rmse-and-its-parameter-derivative}{%
\subsection{S7. Energy loss: reaction RMSE and its parameter
derivative}\label{s7.-energy-loss-reaction-rmse-and-its-parameter-derivative}}

Let \(c_{rm}\) be the coefficient of system \(m\) in reaction \(r\).
Atomization energies use positive atomic coefficients and a negative
molecular coefficient; ionization energies use cation minus neutral.
Define

\[
\Delta E_r(\boldsymbol\theta)=\sum_m c_{rm}E_m(\boldsymbol\theta),
\qquad
\varepsilon_r=\Delta E_r-\Delta E_r^{\mathrm{ref}},
\qquad
\Delta E_r^{\mathrm{ref}}[E_h]
=\frac{\Delta E_r^{\mathrm{ref}}[\mathrm{kcal/mol}]}
{627.5094740631}.
\tag{S54}
\]

The energy loss is the root-mean-square error over energy records,

\[
\mathcal L_E
=\left(\frac1{N_E}\sum_{r=1}^{N_E}\varepsilon_r^2\right)^{1/2}.
\tag{S55}
\]

For \(\mathcal L_E>0\), direct differentiation gives

\[
\frac{\partial\mathcal L_E}{\partial\varepsilon_r}
=\frac{\varepsilon_r}{N_E\mathcal L_E},
\qquad
\frac{\partial\mathcal L_E}{\partial E_m}
=\sum_r\frac{c_{rm}\varepsilon_r}{N_E\mathcal L_E},
\qquad
\frac{\partial\mathcal L_E}{\partial\theta_\ell}
=\sum_m\frac{\partial\mathcal L_E}{\partial E_m}
\frac{\partial E_m}{\partial\theta_\ell}.
\tag{S56}
\]

Thus an atomic anchor contributes to the gradient once for every
reaction in which it appears, with its stoichiometric coefficient. The
energy targets and the coefficients are constants. The code uses an
unweighted mean over the six energy records within \(\mathcal L_E\); the
global energy weight is applied outside the square root. Using MSE,
moving \(w_E\) inside the square root, or separately averaging the AE
and IP subsets would change the training objective.

At an exactly zero residual vector, the norm in Eq. (S55) is not
differentiable. The implementation chooses the zero subgradient by
setting the energy adjoint to zero when \(\mathcal L_E=0\). This
convention avoids a division by zero without introducing an additional
smoothing constant.

\hypertarget{s8.-stationarity-loss-the-squared-predicted-rotation}{%
\subsection{S8. Stationarity loss: the squared predicted
rotation}\label{s8.-stationarity-loss-the-squared-predicted-rotation}}

For the set \(\mathcal G\) of stationarity-training systems, the
implemented loss is

\[
\mathcal L_G
=\sum_{m\in\mathcal G}
\|P_mg_m(\boldsymbol\theta)\|_2^2
=\sum_{m\in\mathcal G}\sum_{\sigma ai}
p_{m,ai\sigma}^2g_{m,ai\sigma}^2
=\sum_{m\in\mathcal G}\|\kappa_{\mathrm{pred},m}\|_2^2.
\tag{S57}
\]

The preconditioner is fixed, so its gradient and parameter derivative
are

\[
\frac{\partial\mathcal L_G}{\partial g_m}
=2P_m^TP_mg_m,\qquad
\frac{\partial\mathcal L_G}{\partial\theta_\ell}
=2\sum_{m\in\mathcal G}
(P_m^TP_mg_m)^T
\frac{\partial g_m}{\partial\theta_\ell}.
\tag{S58}
\]

There is no averaging over systems, occupied--virtual pairs, electron
number, or basis size in this term. The diagonal \(P_m\) is symmetric in
the present implementation, giving the elementwise code expression
\(2p_m^2g_m\).

The loss penalizes an estimated rotation amplitude. It is distinct from
the approximate energy lowering of the quadratic surrogate:

\[
\Delta E_{\mathrm{relax}}^{\mathrm{sur}}
=-\frac12g^TPg,
\qquad
\|\kappa_{\mathrm{pred}}\|^2=g^TP^TPg.
\tag{S59}
\]

The former has one power of the inverse curvature; the implemented G
term has two. If \(\mathcal L_G=0\) and all retained preconditioner
entries are finite and nonzero, the occupied--virtual Fock blocks vanish
for the systems in \(\mathcal G\). This establishes stationarity in
those rotation directions. It does not establish positive orbital
curvature, a global ground state, or exact equality to the correlated
density.

\hypertarget{s9.-density-response-loss-and-its-residual-adjoint}{%
\subsection{S9. Density-response loss and its residual
adjoint}\label{s9.-density-response-loss-and-its-residual-adjoint}}

\hypertarget{s9.1.-definition-and-normalization}{%
\subsubsection{S9.1. Definition and
normalization}\label{s9.1.-definition-and-normalization}}

For one system, the density penalty is

\[
D_m
=\int\left[
\rho_{\mathrm{pred},m}(\mathbf r)
-\rho_m^{\mathrm{CC}}(\mathbf r)\right]^2d\mathbf r
\simeq\sum_\ell w_{m\ell}r_{m\ell}^2
=r_m^TW_mr_m.
\tag{S60}
\]

The training density loss is a mean over the distinct density-training
systems:

\[
\mathcal L_D=\frac1{N_D}\sum_{m\in\mathcal D}D_m,
\qquad N_D=|\mathcal D|.
\tag{S61}
\]

No square root is taken, and no electron-count or grid-point
normalization is applied. Both spin responses are added in Eq. (S51)
before forming \(r_m^2\). Consequently the loss contains the cross term
between the up- and down-spin density errors. Opposite spin-density
changes can cancel in this total-density loss; a sum of separate squared
spin errors would be a different objective.

The grid discretization uses the same quadrature as the fixed orbital
arrays. A stored sampled target is used directly where supplied;
otherwise the target is evaluated from the AO Lambda density on that
grid. Each grid value is weighted by its quadrature volume
\(w_{m\ell}\). Averaging an unweighted pointwise MSE would change both
the spatial measure and its dependence on the grid.

\hypertarget{s9.2.-derivative-with-respect-to-the-orbital-residual}{%
\subsubsection{S9.2. Derivative with respect to the orbital
residual}\label{s9.2.-derivative-with-respect-to-the-orbital-residual}}

Since \(r_m=d_m-B_mP_mg_m\) and \(d_m,B_m,P_m,W_m\) are fixed,

\[
\delta D_m
=2r_m^TW_m\delta r_m
=-2r_m^TW_mB_mP_m\delta g_m,
\qquad
\frac{\partial D_m}{\partial g_m}
=-2P_m^TB_m^TW_mr_m.
\tag{S62}
\]

Using the factor of two in the definition of \(B\), the component
expression is

\[
\frac{\partial D_m}{\partial g_{m,ai\sigma}}
=-4p_{m,ai\sigma}\sum_\ell w_{m\ell}r_{m\ell}
\phi^0_{ma\sigma}(\mathbf r_\ell)
\phi^0_{mi\sigma}(\mathbf r_\ell).
\tag{S63}
\]

The factor \(-4\) combines the derivative of the square, the first-order
orbital-density factor, and the minus sign in \(\kappa=-Pg\). The
density weight and the factor \(1/N_D\) are applied after this
per-system adjoint is constructed. In density\_response
(\nolinkurl{src/ccxc/xcnn_density_response.py}), the quadrature
contraction is implemented as a virtual-orbital matrix multiplied by the
weighted residual and the occupied-orbital matrix, followed by
multiplication by \(-4p\).

For fixed reference arrays the Hessian with respect to the residual
vector is

\[
\frac{\partial^2D_m}{\partial g_m\partial g_m^T}
=2P_m^TB_m^TW_mB_mP_m.
\tag{S64}
\]

It is positive semidefinite when the quadrature weights are nonnegative.
This statement concerns \(g_m\) as an independent variable. It does not
make the objective convex in the nonlinear neural parameters.

\hypertarget{s9.3.-finite-inversion-error-and-the-difference-from-g}{%
\subsubsection{S9.3. Finite inversion error and the difference from
G}\label{s9.3.-finite-inversion-error-and-the-difference-from-g}}

Expanding Eq. (S60) makes the role of the inversion mismatch explicit:

\[
D_m=d_m^TW_md_m
-2g_m^TP_m^TB_m^TW_md_m
+g_m^TP_m^TB_m^TW_mB_mP_mg_m.
\tag{S65}
\]

At \(g_m=0\), \(D_m=d_m^TW_md_m\), which is generally nonzero. The
density adjoint can also be nonzero at that point because it contains
\(B_m^TW_md_m\). Thus the G and D objectives need not have the same
optimum when inversion error is finite. G controls the magnitude of the
predicted orbital rotation, while D controls its real-space consequence
relative to the correlated target. Even with \(d_m=0\), the metric
\(B_m^TW_mB_m\) is generally neither diagonal nor proportional to the
identity, so D is not simply a rescaled G loss.

The per-system diagnostic \texttt{DP} is \(D_m\). \nolinkurl{predicted_L1}
is \(\int|r_m|\,d\mathbf r\), and \nolinkurl{predicted_L2} is
\(\sqrt{D_m}\), with a numerical nonnegative guard in the square root.
The mean of the individual \(L_2\) values is not the square root of the
mean DP. The density loss used for fitting is the mean DP in Eq. (S61),
not either of those alternative summaries.

\hypertarget{s10.-joint-objective-and-exact-blocked-parameter-differentiation}{%
\subsection{S10. Joint objective and exact blocked parameter
differentiation}\label{s10.-joint-objective-and-exact-blocked-parameter-differentiation}}

\hypertarget{s10.1.-residual-adjoints}{%
\subsubsection{S10.1. Residual
adjoints}\label{s10.1.-residual-adjoints}}

The implemented scalar objective is

\[
\mathcal L(\boldsymbol\theta)
=w_E\mathcal L_E+w_G\mathcal L_G+w_D\mathcal L_D.
\tag{S66}
\]

The numerical weights are defined for the atomic-unit conventions above.
\(\mathcal L_E\) has units of \(E_h\), \(\mathcal L_G\) is a squared
dimensionless rotation, and \(\mathcal L_D\) has the
density-squared-volume scale \(a_0^{-3}\). Equal numerical coefficients
therefore do not imply equally normalized or dimensionless
contributions.

The loss depends on the trainable parameters through each fixed-orbital
total energy \(E_m(\boldsymbol\theta)\) and orbital gradient
\(g_m(\boldsymbol\theta)\). Define the weighted energy adjoint

\[
a_m=\frac{\partial\mathcal L}{\partial E_m}
=\frac{w_E}{N_E\mathcal L_E}
\sum_r c_{rm}\varepsilon_r
\quad(\mathcal L_E>0),
\qquad a_m=0\quad(\mathcal L_E=0),
\tag{S67}
\]

and the weighted orbital-gradient adjoint

\[
b_m=\frac{\partial\mathcal L}{\partial g_m}
=2w_G\,\mathbf1_{m\in\mathcal G}P_m^TP_mg_m
-\frac{2w_D}{N_D}\,
\mathbf1_{m\in\mathcal D}P_m^TB_m^TW_mr_m.
\tag{S68}
\]

When \(w_D=0\), the second term is not evaluated and density-target
fields are not required. The code still evaluates the stationarity
diagnostic when \(w_G=0\), but its adjoint contribution is zero. The
full chain rule is

\[
\frac{\partial\mathcal L}{\partial\theta_\ell}
=\sum_m\left[
a_m\frac{\partial E_m}{\partial\theta_\ell}
+b_m^T\frac{\partial g_m}{\partial\theta_\ell}
\right].
\tag{S69}
\]

All model dependence of the Fock residual is retained in this
derivative. No derivative through the precomputed CCSD or WY
calculations is requested by Eq. (S69), because those calculations
define fixed training inputs.

\hypertarget{s10.2.-invariant-directions-for-the-orbital-adjoint}{%
\subsubsection{S10.2. Invariant directions for the orbital
adjoint}\label{s10.2.-invariant-directions-for-the-orbital-adjoint}}

Let \(J^x_{m\ell}\) be the Jacobian of the local raw invariants with
respect to the orbital-rotation vector at the fixed determinant. For an
arbitrary direction \(b_m\), define \(v_{m\ell}=J^x_{m\ell}b_m\). The
first two entries are the density variations in Eq. (S43) evaluated with
\(\kappa=b_m\). Their gradient-invariant entries are

\[
\begin{aligned}
v_{n_\sigma}&=\delta_b n_\sigma,\\
v_{\sigma_{\uparrow\uparrow}}
&=2\nabla n_\uparrow^0\cdot\nabla\delta_b n_\uparrow,\\
v_{\sigma_{\uparrow\downarrow}}
&=\nabla n_\downarrow^0\cdot\nabla\delta_b n_\uparrow
+\nabla n_\uparrow^0\cdot\nabla\delta_b n_\downarrow,\\
v_{\sigma_{\downarrow\downarrow}}
&=2\nabla n_\downarrow^0\cdot\nabla\delta_b n_\downarrow,\\
v_{\tau_\sigma}
&=\sum_{ai}\nabla\phi^0_{a\sigma}\cdot
\nabla\phi^0_{i\sigma}\,b_{m,ai\sigma}.
\end{aligned}
\tag{S70}
\]

The final line is omitted for GGA. This direction is a contraction
needed for the gradient calculation; it is not an additional physical
orbital update. The functions named \texttt{direction} in
\nolinkurl{src/ccxc/xcnn_tensor_orbitals.py} and
\nolinkurl{src/ccxc/mgga_projection.py} implement Eq. (S70).

The neural part of the orbital gradient follows from differentiation of
the quadrature energy:

\[
g_m^{\mathrm{NN}}
=\sum_\ell w_{m\ell}(J^x_{m\ell})^T
\nabla_{\mathbf x}q_{\boldsymbol\theta}(\mathbf x_{m\ell}),
\qquad
b_m^Tg_m^{\mathrm{NN}}
=\sum_\ell w_{m\ell}v_{m\ell}^T
\nabla_{\mathbf x}q_{\boldsymbol\theta}(\mathbf x_{m\ell}).
\tag{S71}
\]

This avoids storing the full Jacobian of every orbital-gradient
component with respect to every network parameter.

\hypertarget{s10.3.-scalar-whose-backward-pass-gives-the-exact-loss-gradient}{%
\subsubsection{S10.3. Scalar whose backward pass gives the exact loss
gradient}\label{s10.3.-scalar-whose-backward-pass-gives-the-exact-loss-gradient}}

After a forward evaluation has supplied \(a_m\) and \(b_m\), hold them
fixed and form the auxiliary scalar

\[
\begin{aligned}
\Phi(\boldsymbol\theta;a,b)
=\sum_m\bigg\{&
\alpha\left[
a_m E^{\mathrm{base}}_{xc,m}
+b_m^Tg_m^{\mathrm{base}}\right]\\
&+\sum_\ell w_{m\ell}
\left[
a_mq_{\boldsymbol\theta}(\mathbf x_{m\ell})
+v_{m\ell}^T\nabla_{\mathbf x}
q_{\boldsymbol\theta}(\mathbf x_{m\ell})
\right]\bigg\}.
\end{aligned}
\tag{S72}
\]

The parameter-independent \(E_m^{(0)}\) and \(g_m^{(0)}\) need not
appear in this scalar. Equations (S47), (S69), and (S71) give

\[
\nabla_{\boldsymbol\theta}\mathcal L
=\left.
\nabla_{\boldsymbol\theta}\Phi(\boldsymbol\theta;a,b)
\right|_{a,b,v\ \mathrm{held\ fixed}}.
\tag{S73}
\]

Holding the residual adjoints fixed is the usual vector--Jacobian
product rule. Differentiating \(a_m\) or \(b_m\) again inside \(\Phi\)
would add terms not present in the first derivative in Eq. (S69). This
operation does not discard the required derivative of \(g_m\) with
respect to the network parameters.

Within each grid block, the implementation creates a local invariant
tensor with gradients enabled, evaluates \(q_{\boldsymbol\theta}\), and
obtains \(\nabla_{\mathbf x}q_{\boldsymbol\theta}\) with
\nolinkurl{create_graph=True}. Backward differentiation of Eq. (S72) then
includes

\[
\frac{\partial}{\partial\theta_\ell}
\nabla_{\mathbf x}q_{\boldsymbol\theta}
=\nabla_{\mathbf x}
\frac{\partial q_{\boldsymbol\theta}}{\partial\theta_\ell}
\tag{S74}
\]

wherever the selected numerical branch is differentiable. These mixed
derivatives are needed because the loss contains a potential/Fock
residual, not only an energy. Disabling the graph for the invariant
derivative would remove the G and D parameter-gradient contributions.

Because Eq. (S72) is a sum over grid blocks, parameter gradients can be
accumulated block by block and the local graph released after each
block. The full-grid residuals and their adjoints are computed before
this backward pass. This is an exact blocking of the stated discrete
objective, apart from floating-point summation order, rather than a
stochastic minibatch approximation.

\hypertarget{s10.4.-mixing-parameter-and-network-gradients}{%
\subsubsection{S10.4. Mixing-parameter and network
gradients}\label{s10.4.-mixing-parameter-and-network-gradients}}

The scalar base-functional parameter has the particularly simple
derivative

\[
\frac{\partial\mathcal L}{\partial\alpha}
=\sum_m\left[
a_mE^{\mathrm{base}}_{xc,m}
+b_m^Tg_m^{\mathrm{base}}\right].
\tag{S75}
\]

Since \(q_{\boldsymbol\theta}=\beta n f_{\boldsymbol\omega}\),

\[
\frac{\partial\mathcal L}{\partial\beta}
=\sum_{m\ell}w_{m\ell}
\left[
a_m n_{m\ell}f_{\boldsymbol\omega}(\mathbf d_{m\ell})
+v_{m\ell}^T\nabla_{\mathbf x}
\bigl(nf_{\boldsymbol\omega}(\mathbf d(\mathbf x))\bigr)
\big|_{\mathbf x_{m\ell}}
\right],
\tag{S76}
\]

\[
\frac{\partial\mathcal L}{\partial\omega_j}
=\sum_{m\ell}w_{m\ell}
\left[
a_m\frac{\partial q_{\boldsymbol\theta}}{\partial\omega_j}
+v_{m\ell}^T\nabla_{\mathbf x}
\frac{\partial q_{\boldsymbol\theta}}{\partial\omega_j}
\right].
\tag{S77}
\]

At initialization, \(\alpha=1\) and \(\beta=0\), so the model is exactly
the selected base functional within its numerical conventions. The
network-weight derivatives in Eq. (S77) initially vanish because they
are proportional to \(\beta\), while the \(\beta\) derivative need not
vanish. The first parameter update can therefore activate the neural
correction even though the initial energy is PBE or SCAN. Subsequent
updates propagate gradients through the network as well as through both
scalar mixing parameters.

\hypertarget{s11.-discrete-implementation-and-training-protocol}{%
\subsection{S11. Discrete implementation and training
protocol}\label{s11.-discrete-implementation-and-training-protocol}}

\hypertarget{s11.1.-cached-arrays-and-their-meaning}{%
\subsubsection{S11.1. Cached arrays and their
meaning}\label{s11.1.-cached-arrays-and-their-meaning}}

For a system with \(N_q\) grid points and \(N_{\mathrm{MO}}\) orbitals
per spin block, define
\(N_{\mathrm{rot}}=\sum_\sigma n_{o\sigma}(N_{\mathrm{MO}}-n_{o\sigma})\).
The following array conventions connect the derivation to the
implementation.

\begin{longtable}[]{@{}
  >{\raggedright\arraybackslash}p{(\columnwidth - 4\tabcolsep) * \real{0.3333}}
  >{\raggedright\arraybackslash}p{(\columnwidth - 4\tabcolsep) * \real{0.3333}}
  >{\raggedright\arraybackslash}p{(\columnwidth - 4\tabcolsep) * \real{0.3333}}@{}}
\toprule\noalign{}
\begin{minipage}[b]{\linewidth}\raggedright
Stored field
\end{minipage} & \begin{minipage}[b]{\linewidth}\raggedright
Shape
\end{minipage} & \begin{minipage}[b]{\linewidth}\raggedright
Meaning
\end{minipage} \\
\midrule\noalign{}
\endhead
\bottomrule\noalign{}
\endlastfoot
\texttt{rho}, GGA & \((2,4,N_q)\) & \(n_\sigma\) and its three Cartesian
derivatives \\
\texttt{rho}, mGGA & \((2,5,N_q)\) & The GGA fields followed by
\(\tau_\sigma\) \\
\texttt{phi} & \((2,4,N_q,N_{\mathrm{MO}})\) & Orbital values and three
Cartesian orbital derivatives \\
\texttt{x} & \((N_q,5)\) or \((N_q,7)\) & Raw invariants in Eq. (S32) \\
\texttt{weights} & \((N_q,)\) & Quadrature weights \\
\texttt{nocc} & \((2,)\) & Occupied-orbital counts in the two spin
blocks \\
\texttt{precond} & \((N_{\mathrm{rot}},)\) & Diagonal entries of \(P\)
in Eq. (S50) \\
\texttt{g0} & \((N_{\mathrm{rot}},)\) & The physical non-XC residual in
Eq. (S46) \\
\texttt{e0} & Scalar & The physical non-XC energy in Eq. (S46) \\
\texttt{epbe}/\texttt{escan} & Scalar & Fixed-orbital base XC energy \\
\texttt{gpbe}/\texttt{gscan} & \((N_{\mathrm{rot}},)\) & Fixed-orbital
base XC projected gradient \\
\nolinkurl{rho_target} & \((N_q,)\) & Total correlated density on the same
grid \\
\end{longtable}

Orbital-gradient vectors concatenate the up-spin and down-spin
virtual-by-occupied matrices. Each matrix is flattened in row-major
order, so the occupied index varies fastest within a virtual row. The
same ordering is used for \(g\), \(P\), \(\kappa\), and the adjoint
\(b\). Empty occupied or virtual blocks contribute zero terms.

The grid cache is built from AO functions and their derivatives, the
fixed WY coefficients and eigenvalues, and the correlated target.
\nolinkurl{src/ccxc/reproduce.py} loads the compact inputs and reconstructs
the cache. Its GGA point builder
(\nolinkurl{src/ccxc/xcnn_fixed_orbitals.py}) constructs Eq. (S46), the
PBE columns and the reference gaps. The mGGA point builder
(\nolinkurl{src/ccxc/xcnn_scan_orbitals.py}) adds \(\tau_\sigma\),
reconstructs the SCAN columns, and retains the same orbital/gap
convention.

\hypertarget{s11.2.-one-parameter-update}{%
\subsubsection{S11.2. One parameter
update}\label{s11.2.-one-parameter-update}}

The implementation first evaluates all molecular energies and orbital
gradients using the current model. Reaction errors, the stationarity
sum, and the mean density penalty are then assembled. Their residual
adjoints are computed from Eqs. (S67) and (S68), and the blocked
backward pass evaluates Eq. (S73). Only after that full-data gradient is
available is one optimizer step taken.

The following pseudocode summarizes that update. The symbols used for
the fields are those in the derivation; the table above gives their
actual cache names. Each backward block contains only the valid neural
grid points with \(n>10^{-20}\).

\begin{Shaded}
\begin{Highlighting}[]
\NormalTok{E, g }\OperatorTok{=}\NormalTok{ evaluate\_all\_fixed\_orbital\_points(model, points)}
\NormalTok{errors }\OperatorTok{=}\NormalTok{ stoichiometry }\OperatorTok{@}\NormalTok{ E }\OperatorTok{{-}}\NormalTok{ reference\_energies\_Eh}
\NormalTok{LE }\OperatorTok{=}\NormalTok{ sqrt(mean(errors}\OperatorTok{**}\DecValTok{2}\NormalTok{))}
\NormalTok{LG }\OperatorTok{=} \BuiltInTok{sum}\NormalTok{(norm(P[m] }\OperatorTok{*}\NormalTok{ g[m])}\OperatorTok{**}\DecValTok{2} \ControlFlowTok{for}\NormalTok{ m }\KeywordTok{in}\NormalTok{ G\_set)}
\NormalTok{residual }\OperatorTok{=}\NormalTok{ \{m: rho0[m] }\OperatorTok{{-}}\NormalTok{ rhoCC[m] }\OperatorTok{{-}}\NormalTok{ B[m] }\OperatorTok{@}\NormalTok{ (P[m] }\OperatorTok{*}\NormalTok{ g[m])}
            \ControlFlowTok{for}\NormalTok{ m }\KeywordTok{in}\NormalTok{ D\_set\}}
\NormalTok{LD }\OperatorTok{=}\NormalTok{ mean(dot(weights[m], residual[m]}\OperatorTok{**}\DecValTok{2}\NormalTok{) }\ControlFlowTok{for}\NormalTok{ m }\KeywordTok{in}\NormalTok{ D\_set)}
\NormalTok{loss }\OperatorTok{=}\NormalTok{ wE }\OperatorTok{*}\NormalTok{ LE }\OperatorTok{+}\NormalTok{ wG }\OperatorTok{*}\NormalTok{ LG }\OperatorTok{+}\NormalTok{ wD }\OperatorTok{*}\NormalTok{ LD}

\NormalTok{a, b }\OperatorTok{=}\NormalTok{ residual\_adjoints(errors, LE, g, residual)  }\CommentTok{\# Eqs. (S67), (S68)}
\NormalTok{a, b }\OperatorTok{=}\NormalTok{ stop\_gradient(a), stop\_gradient(b)}
\NormalTok{model.zero\_grad(set\_to\_none}\OperatorTok{=}\VariableTok{True}\NormalTok{)}
\ControlFlowTok{for}\NormalTok{ m }\KeywordTok{in}\NormalTok{ points:}
\NormalTok{    backward(alpha }\OperatorTok{*}\NormalTok{ (a[m] }\OperatorTok{*}\NormalTok{ Ebase[m] }\OperatorTok{+}\NormalTok{ dot(b[m], gbase[m])))}
    \ControlFlowTok{for}\NormalTok{ block }\KeywordTok{in}\NormalTok{ valid\_grid\_blocks(m):}
\NormalTok{        x }\OperatorTok{=}\NormalTok{ fixed\_invariants(block).requires\_grad\_(}\VariableTok{True}\NormalTok{)}
\NormalTok{        q }\OperatorTok{=}\NormalTok{ model.correction(x)}
\NormalTok{        qx }\OperatorTok{=}\NormalTok{ autograd\_grad(q.}\BuiltInTok{sum}\NormalTok{(), x, create\_graph}\OperatorTok{=}\VariableTok{True}\NormalTok{)}
\NormalTok{        v }\OperatorTok{=}\NormalTok{ invariant\_direction(m, b[m], block)    }\CommentTok{\# Eq. (S70)}
\NormalTok{        backward(}\BuiltInTok{sum}\NormalTok{(block.weights }\OperatorTok{*}\NormalTok{ (a[m] }\OperatorTok{*}\NormalTok{ q }\OperatorTok{+}\NormalTok{ row\_dot(v, qx))))}
\NormalTok{optimizer.step()}
\end{Highlighting}
\end{Shaded}

When the density weight is zero, the residual and LD operations are
skipped. This pseudocode does not differentiate through a new SCF at
each update. The backward pass differentiates the parameter-dependent XC
correction and Fock residual evaluated on fixed orbitals.

All six configurations use \texttt{torch.optim.RAdam} with learning rate
\(10^{-4}\) and zero weight decay. The PyTorch 2.5.1 defaults are first-
and second-moment coefficients \((0.9,0.999)\) and \(\epsilon=10^{-8}\);
the RAdam rectification and update are delegated to that implementation.
Network initialization uses seed 0 and the ordinary PyTorch linear-layer
initialization. The constructors create the linear layers at the default
floating-point dtype and then convert them to float64. Reproducing that
initialization requires preserving the default float32 construction
convention before the conversion. Both scalar mixing parameters are
explicitly float64.

The training block size is 32768 grid points. Each update uses all six
energy records and all 11 stationarity/density systems. Model and
optimizer checkpoints are written every 50 completed updates and at the
requested endpoint. A checkpoint at update \(s\) contains the parameters
after \(s\) updates, before the transition to \(s+1\). No network-weight
penalty is added.

\hypertarget{s11.3.-data-sets-weights-and-selected-checkpoints}{%
\subsubsection{S11.3. Data sets, weights, and selected
checkpoints}\label{s11.3.-data-sets-weights-and-selected-checkpoints}}

The energy set has four atomization records, H\(_2\), LiH, O\(_2\), and
CO, and two ionization records, O and Ne. The 13 orbital systems are H,
Li, C, O, Ne, H\(_2\), LiH, Li\(_2\), B\(_2\), O\(_2\), CO, O\(^+\), and
Ne\(^+\). The 11 G/D systems exclude the C and O atomic energy anchors.
C and O still contribute to the energy loss through their reaction
coefficients.

\begin{longtable}[]{@{}
  >{\raggedright\arraybackslash}p{(\columnwidth - 8\tabcolsep) * \real{0.1875}}
  >{\raggedright\arraybackslash}p{(\columnwidth - 8\tabcolsep) * \real{0.1875}}
  >{\raggedright\arraybackslash}p{(\columnwidth - 8\tabcolsep) * \real{0.1875}}
  >{\raggedleft\arraybackslash}p{(\columnwidth - 8\tabcolsep) * \real{0.2500}}
  >{\raggedright\arraybackslash}p{(\columnwidth - 8\tabcolsep) * \real{0.1875}}@{}}
\toprule\noalign{}
\begin{minipage}[b]{\linewidth}\raggedright
Family
\end{minipage} & \begin{minipage}[b]{\linewidth}\raggedright
Loss scheme
\end{minipage} & \begin{minipage}[b]{\linewidth}\raggedright
\((w_E,w_G,w_D)\)
\end{minipage} & \begin{minipage}[b]{\linewidth}\raggedleft
Selected update
\end{minipage} & \begin{minipage}[b]{\linewidth}\raggedright
Fixed-input branch
\end{minipage} \\
\midrule\noalign{}
\endhead
\bottomrule\noalign{}
\endlastfoot
NN-GGA & E & \((50,0,0)\) & 910 & \nolinkurl{data/original/} \\
NN-GGA & EG & \((50,50,0)\) & 2640 & \nolinkurl{data/original/} \\
NN-GGA & EGD & \((50,50,500)\) & 2280 & \nolinkurl{data/original/} \\
NN-mGGA & E & \((50,0,0)\) & 465 & \nolinkurl{data/original/} \\
NN-mGGA & EG & \((50,50,0)\) & 1590 & \nolinkurl{data/original/} \\
NN-mGGA & EGD & \((50,50,200)\) & 1500 &
\nolinkurl{data/current_egd200/} \\
\end{longtable}

The exact geometries, spin/charge assignments, stoichiometric
coefficients, and target energies are in the two original
(\nolinkurl{data/original/inputs.json}) and current EGD200
(\nolinkurl{data/current_egd200/inputs.json}) manifests. The mGGA EGD200
campaign uses revised inputs and energy targets; its comparison with the
original mGGA E/EG runs is therefore not a strict same-input ablation of
\(w_D\) alone. The configuration files preserve those distinctions.

The training objective is separate from the checkpoint-selection score.
NN-GGA selection used validation energy RMSE in kcal/mol plus \(5000\)
times raw validation DP; the selected mGGA models used the corresponding
\(1000\)-DP score. These validation scores are not substituted into Eq.
(S66). The selection cohorts, scan resolutions and incomplete historical
validation points are documented in \nolinkurl{docs/reproduction.md} and
\nolinkurl{provenance/checkpoint_selection.json}.

\hypertarget{s11.4.-function-to-equation-map}{%
\subsubsection{S11.4. Function-to-equation
map}\label{s11.4.-function-to-equation-map}}

\begin{longtable}[]{@{}
  >{\raggedright\arraybackslash}p{(\columnwidth - 4\tabcolsep) * \real{0.3333}}
  >{\raggedright\arraybackslash}p{(\columnwidth - 4\tabcolsep) * \real{0.3333}}
  >{\raggedright\arraybackslash}p{(\columnwidth - 4\tabcolsep) * \real{0.3333}}@{}}
\toprule\noalign{}
\begin{minipage}[b]{\linewidth}\raggedright
Operation
\end{minipage} & \begin{minipage}[b]{\linewidth}\raggedright
Repository implementation
\end{minipage} & \begin{minipage}[b]{\linewidth}\raggedright
Equations
\end{minipage} \\
\midrule\noalign{}
\endhead
\bottomrule\noalign{}
\endlastfoot
UHF, UCCSD, Lambda and AO density &
\nolinkurl{scripts/prepare_references.py}, using PySCF UCCSD RDM routines
& (S3)--(S20) \\
Regularized WY objective, gradient and Hessian &
\nolinkurl{src/ccxc/wy_polish.py}, \nolinkurl{_Objective} & (S24)--(S30) \\
Potential and AO coordinate transformations &
\nolinkurl{src/ccxc/wy_polish.py}, \texttt{invert} & (S23), (S29),
(S30) \\
Neural GGA and PySCF derivatives & \nolinkurl{src/ccxc/kasim_nn.py} &
(S31)--(S39) \\
Neural mGGA and PySCF derivatives & \nolinkurl{src/ccxc/xcnn_scan.py} &
(S31)--(S39) \\
Fixed-orbital GGA arrays and projection &
\nolinkurl{src/ccxc/xcnn_fixed_orbitals.py},
\nolinkurl{src/ccxc/xcnn_tensor_orbitals.py} & (S42), (S45)--(S47) \\
Tau projection and invariant direction &
\nolinkurl{src/ccxc/mgga_projection.py},
\nolinkurl{src/ccxc/xcnn_scan_orbitals.py} & (S44), (S45), (S70) \\
Predicted density and density adjoint &
\nolinkurl{src/ccxc/xcnn_density_response.py},
\nolinkurl{response_density}, \nolinkurl{density_response} & (S50)--(S53),
(S60)--(S65) \\
GGA E/G/D assembly & \nolinkurl{src/ccxc/xcnn_density_response.py},
\nolinkurl{loss_and_gradient} & (S54)--(S69) \\
mGGA E/G/D assembly & \nolinkurl{src/ccxc/xcnn_scan_response.py},
\nolinkurl{loss_and_gradient} & (S54)--(S69) \\
Blocked parameter backward pass &
\nolinkurl{src/ccxc/xcnn_tensor_orbitals.py},
\nolinkurl{src/ccxc/xcnn_scan_orbitals.py}, \nolinkurl{backward_point} &
(S70)--(S77) \\
Optimizer and checkpoint loop & \nolinkurl{scripts/train.py} & Section
S11.2 \\
\end{longtable}

\hypertarget{s12.-verification-and-scope-of-the-approximation}{%
\subsection{S12. Verification and scope of the
approximation}\label{s12.-verification-and-scope-of-the-approximation}}

The algebra in this supplement was checked independently on small
numerical systems. The explicit unsymmetrized CCSD density in Eqs.
(S12)--(S15) was compared with direct creation/annihilation-operator
evaluation of Eq. (S11) in a five-spin-orbital Fock space with two
electrons and random antisymmetric amplitudes. The largest matrix
difference was \(1.11\times10^{-16}\), with unit left/right overlap and
conserved particle number. This verifies the tensor contractions and
index convention, without treating the random amplitudes as a physical
converged CCSD state.

The WY gradient and Hessian were compared with central differences of
the matrix eigenvalue objective for occupation factors one and two. The
raw/whitened gradient transformation was also checked. The largest
absolute gradient and Hessian differences were \(7.23\times10^{-11}\)
and \(1.63\times10^{-11}\), respectively; the coordinate-transform
difference was below \(3.2\times10^{-16}\). The combined energy and
orbital-residual adjoints in Eqs. (S67) and (S68) were checked against
finite differences of the discrete E/G/D objective. The corresponding
largest differences were \(2.72\times10^{-10}\) and
\(1.20\times10^{-8}\). These are local algebra checks, not additional
production electronic-structure results.

The released density-response tests
(\nolinkurl{tests/test_xcnn_density_response.py}) check the response
sign and factor against explicit Fock diagonalization, the density
adjoint against autograd, particle-number conservation, spin
cancellation, zero-D compatibility, and parameter derivatives. The mGGA
tests (\nolinkurl{tests/test_xcnn_scan.py}) check invariant derivatives,
the SCAN limit, GKS matrix consistency, tau orbital directions, and the
joint parameter gradient. The original release recorded 16 passing tests
and one CUDA test skipped on that host. It also recorded all six
full-training-input initial-loss/gradient/first-RAdam checks as
accepted; these results are in \nolinkurl{provenance/release_checks.json}.
They are existing release evidence, not a claim that full training was
repeated when this supplement was added. The new algebra-check record is
\nolinkurl{provenance/sm_derivation_checks.json}.

The variational identities for the fixed-orbital energy and its
derivatives do not remove the response approximation in Eq. (S50). A
small G loss, a small predicted-density D loss, or a converged WY
optimizer does not by itself establish the quality or stability of a
self-consistent solution of the learned functional. That solution
requires a separate SCF using the full candidate GKS operator. The final
published benchmark values concern such separate calculations and are
stored with their own acceptance scope.

For evaluation, DP99 is a mean over 99 density records corresponding to
98 distinct molecules, including the duplicated ethylene-oxide record.
That record weighting is distinct from the mean over 11 distinct
training systems in Eq. (S61). The current AE103/DP99 evaluation
definitions preserve the corrected CH\(_3\)CN/N\(_2\)H\(_4\) geometries
and C\(_2\)H Sigma reference assignment. They do not overwrite the fixed
inputs of a historical trained checkpoint. The released mGGA EGD1500
benchmark did not include stability certification. These protocol
distinctions are part of interpreting the model results, rather than
changes to the E/G/D formulas.

\hypertarget{s13.-linear-hybrid-fitting-cdf26d}{%
\subsection{S13. Linear hybrid fitting
(CDF26D)}\label{s13.-linear-hybrid-fitting-cdf26d}}

\hypertarget{s13.1.-functional-form-and-fixed-orbital-energy}{%
\subsubsection{S13.1. Functional form and fixed-orbital
energy}\label{s13.1.-functional-form-and-fixed-orbital-energy}}

CDF26D has the MN15 functional form {[}9{]} with 59 fitted parameters:
40 semilocal exchange coefficients, 18 correlation coefficients, and the
global exact-exchange fraction \(a_{\mathrm{HF}}\). Write the 58
semilocal coefficients as \(\boldsymbol\theta_{58}\) and
\(\boldsymbol\theta=(\boldsymbol\theta_{58},a_{\mathrm{HF}})\). For a
fixed determinant with spin density matrices \(D_\sigma\), the energy is
affine in the parameters:

\[
E(\boldsymbol\theta)=e_0+\mathbf E_{58}^T\boldsymbol\theta_{58}+a_{\mathrm{HF}}E_{\mathrm{HF}},
\qquad
E_{\mathrm{HF}}=-\frac12\sum_\sigma\operatorname{Tr}(K_\sigma D_\sigma).
\tag{S78}
\]

Here \(e_0\) contains the nuclear-repulsion, one-electron, and Hartree
energies, and the \(k\)th entry of \(\mathbf E_{58}\) is the quadrature
energy of the \(k\)th semilocal basis term evaluated on the fixed
orbitals. The 59th column is the exact-exchange energy. A reaction
energy is a fixed linear combination of system energies, so each
reaction residual is an affine function
\((A_d\boldsymbol\theta+\mathbf b_d)_i\) of the parameters, with \(A_d\)
and \(\mathbf b_d\) in hartree.

\hypertarget{s13.2.-brillouin-residual-of-a-hybrid-and-its-linearity}{%
\subsubsection{S13.2. Brillouin residual of a hybrid and its
linearity}\label{s13.2.-brillouin-residual-of-a-hybrid-and-its-linearity}}

The generalized Kohn--Sham matrix of the parametrized hybrid is

\[
F_\sigma(\boldsymbol\theta)=h+J[D]+\sum_{k=1}^{58}\theta_kV^{(k)}_\sigma-a_{\mathrm{HF}}K_\sigma,
\tag{S79}
\]

where \(V^{(k)}_\sigma\) is the AO matrix of the functional derivative
of the \(k\)th semilocal term, built as in Eq. (S37) including its
kinetic-density part. On the fixed WY orbitals of a density-constraint
system, the occupied--virtual block is therefore

\[
F^\sigma_{ai}(\boldsymbol\theta)
=f^{(0)}_{ai\sigma}+\sum_{k=1}^{58}\theta_k\,C^{0T}_{a\sigma}V^{(k)}_\sigma C^0_{i\sigma}
-a_{\mathrm{HF}}\,C^{0T}_{a\sigma}K_\sigma C^0_{i\sigma},
\qquad
f^{(0)}_{ai\sigma}=C^{0T}_{a\sigma}(h+J[D^0])C^0_{i\sigma}.
\tag{S80}
\]

Equation (S80) is affine in \(\boldsymbol\theta\), including the
exact-exchange column. Through Eqs. (S50), (S51), and (S53), the
predicted rotation, the predicted density, and the density residual
\(r_m=d_m-B_mP_mg_m\) are then affine as well:

\[
r_m(\boldsymbol\theta)=r_m^{(0)}+R_m^{(1)}\boldsymbol\theta .
\tag{S81}
\]

Because exact exchange is nonlocal, its occupied--virtual block need not
be representable by any local potential, and \(F_{ai}=0\) need not be
attainable on local-potential (WY) orbitals for all systems
simultaneously. The density information is therefore imposed as an upper
bound on the predicted density error (Section S13.4), not as a penalty
to be driven to zero.

\hypertarget{s13.3.-normalized-energy-objective}{%
\subsubsection{S13.3. Normalized energy
objective}\label{s13.3.-normalized-energy-objective}}

Let \(M=25\) be the number of training data sets, \(N_d\) the number of
records in set \(d\), and \(E^{\mathrm{std}}_d\) the GSCDB137 error
scale of set \(d\) {[}10{]} (Section S14.3). The energy objective is the
set-balanced mean of squared normalized residuals,

\[
\mathcal L_E^{\mathrm{lin}}(\boldsymbol\theta)
=\frac1M\sum_{d=1}^{M}\frac1{N_d}\sum_{i=1}^{N_d}
\left[\frac{627.5094740631\,(A_d\boldsymbol\theta+\mathbf b_d)_i}{E^{\mathrm{std}}_d}\right]^2
=\|B\boldsymbol\theta+\mathbf c\|_2^2,
\tag{S82}
\]

obtained by scaling each row of \((A_d,\mathbf b_d)\) by
\(627.5094740631/(E^{\mathrm{std}}_d\sqrt{MN_d})\). Every data set has
equal total weight, and every record within a set has equal weight, so
large sets cannot dominate the fit by size alone. The reported NER
(Section S14.3) uses absolute rather than squared normalized errors; Eq.
(S82) is its smooth quadratic counterpart.

\hypertarget{s13.4.-constraints}{%
\subsubsection{S13.4. Constraints}\label{s13.4.-constraints}}

The fit is

\[
\min_{\boldsymbol\theta}\ \|B\boldsymbol\theta+\mathbf c\|_2^2
\quad\text{subject to}\quad
\|D_s\boldsymbol\theta\|_2^2\le 1600,\qquad
\left\|R\begin{bmatrix}1\\\boldsymbol\theta\end{bmatrix}\right\|_2^2\le 3\times10^{-4},\qquad
0\le a_{\mathrm{HF}}\le 1 .
\tag{S83}
\]

\emph{Smoothness.} \(D_s\) contains the 46 differences between adjacent
semilocal coefficients used for CF22D {[}11{]}; its exact-exchange
column is zero. The bound limits oscillation of the enhancement factors.

\emph{First-order density constraint.} Stack the quadrature-weighted
residuals of Eq. (S81) for all DP99 records, with the record weights of
Section S14.2 and the mean over records of Eq. (S61):

\[
\mathcal L_D^{\mathrm{DP99}}(\boldsymbol\theta)
=\frac{1}{\sum_m\omega_m}\sum_m\omega_m\,r_m^TW_mr_m
=\left\|\big[\,\mathbf a_0\ \ A_1\,\big]\begin{bmatrix}1\\\boldsymbol\theta\end{bmatrix}\right\|_2^2 .
\tag{S84}
\]

A thin QR factorization \([\mathbf a_0\ A_1]=QR\) compresses the stacked
matrix to the \(60\times60\) upper-triangular factor \(R\) without
changing the quadratic form, since \(Q^TQ=I\). The bound
\(3\times10^{-4}\) corresponds to a predicted DP99 of
\(0.3\times10^{-3}\) a.u. No separate stationarity term \(\mathcal L_G\)
is used for the hybrid; the Brillouin residual enters through the
predicted density. The density term carries no weight in the objective;
it enters only through the bound.

\hypertarget{s13.5.-convexity-and-the-global-optimum-of-each-fit}{%
\subsubsection{S13.5. Convexity and the global optimum of each
fit}\label{s13.5.-convexity-and-the-global-optimum-of-each-fit}}

For fixed design matrices, (i) the objective has Hessian
\(2B^TB\succeq0\); (ii) the smoothness constraint is a convex quadratic
with matrix \(D_s^TD_s\succeq0\); (iii) the density constraint is the
squared norm of an affine function of \(\boldsymbol\theta\), with
quadratic part \(R_1^TR_1\succeq0\), where \(R=[\mathbf r_0\ R_1]\); and
(iv) the exact-exchange interval is a convex set. Problem (S83) is
therefore a convex quadratically constrained least-squares problem. For
a convex problem, any point satisfying the Karush--Kuhn--Tucker (KKT)
conditions is a global minimizer.

The solver introduces multipliers \(\lambda,\mu\ge0\) for the two
quadratic constraints. For given multipliers the stationarity condition
is the regularized least-squares problem

\[
\min_{\boldsymbol\theta}\left\|
\begin{bmatrix}B\\ \sqrt{\lambda}\,D_s\\ \sqrt{\mu}\,R_1\end{bmatrix}\boldsymbol\theta
+\begin{bmatrix}\mathbf c\\ \mathbf 0\\ \sqrt{\mu}\,\mathbf r_0\end{bmatrix}
\right\|_2^2 ,
\tag{S85}
\]

solved by singular-value decomposition (LAPACK \texttt{gelsd}). The
active set is determined by testing the unconstrained solution, each
single active constraint, and both constraints active, adjusting
\(\lambda\) and \(\mu\) until the active constraints hold with equality.
If \(a_{\mathrm{HF}}\) leaves \([0,1]\), the problem is re-solved with
\(a_{\mathrm{HF}}\) fixed at the violated bound, and the sign of the
bound multiplier is checked. The accepted solution satisfies primal
feasibility, dual feasibility, KKT stationarity, and complementary
slackness, with a vanishing duality gap. No learning rate, epoch count,
or loss weight enters; \(\lambda\) and \(\mu\) are solved for. The
certificate applies to the fit with fixed matrices; it does not assert
uniqueness of the parameters.

\hypertarget{s13.6.-self-consistent-cycle}{%
\subsubsection{S13.6. Self-consistent
cycle}\label{s13.6.-self-consistent-cycle}}

The energy design matrices \(A_d,\mathbf b_d\) are evaluated on the
self-consistent orbitals of the current parameter set. After each convex
fit, all training systems are recomputed self-consistently with the new
parameters, and the matrices are rebuilt. The density-constraint
matrices are built on the fixed WY orbitals of the DP99 CCSD densities
and do not change between cycles. The self-consistent calculations use
unrestricted global-hybrid GKS with direct (non-density-fitted) Coulomb
and exchange, PySCF grid level 6 {[}7{]}, and the basis sets listed in
Tables S2 and S3. For the final parameter set, 1718 energy calculations
and 98 density calculations were completed. The final exact-exchange
fraction is \(a_{\mathrm{HF}}=0.488709463180\).

\hypertarget{s13.7.-dispersion}{%
\subsubsection{S13.7. Dispersion}\label{s13.7.-dispersion}}

D4 dispersion {[}12{]} is added to the fixed XC functional with native
EEQ charges determined by geometry and total charge, including the
three-body ATM term (\(s_9=1\)), with \(s_6=1\) and \(\alpha=16\) fixed.
The damping parameters \(p=(s_8,a_1,a_2)\) minimize the set-balanced
mean NER over the 25 training sets,

\[
\mathcal L_{\mathrm{D4}}(p)=\frac1{25}\sum_{d\in\mathrm{train}}\frac1{N_dE^{\mathrm{std}}_d}\sum_i
\left|\Delta E^{\mathrm{SCF}}_{d,i}+\Delta E^{\mathrm{D4}}_{d,i}(p)-\Delta E^{\mathrm{ref}}_{d,i}\right|
\tag{S86}
\]

(energies in kcal/mol), with the 59 XC parameters fixed and no further
SCF. Because the damping parameters enter nonlinearly, this step uses
differential evolution (seed 27, population size 8, at most 100
generations, tolerance \(10^{-8}\)) over \(s_8\in[0,20]\),
\(a_1\in[0,1.5]\), \(a_2\in[1,20]\), followed by Powell refinement from
three starting points (at most 1200 function evaluations each,
\(x_{\mathrm{tol}}=10^{-7}\), \(f_{\mathrm{tol}}=10^{-10}\)). The test
sets play no role in this selection. The final values are
\(s_8=19.99995316\), \(a_1=0.5535564521\), and \(a_2=11.35437340\).
Dispersion added at fixed geometry does not change orbitals or
densities, so all density metrics of CDF26D are those of the XC
functional.

\hypertarget{s14.-data-sets-and-evaluation-metrics}{%
\subsection{S14. Data sets and evaluation
metrics}\label{s14.-data-sets-and-evaluation-metrics}}

\hypertarget{s14.1.-neural-functional-training-and-validation-systems}{%
\subsubsection{S14.1. Neural-functional training and validation
systems}\label{s14.1.-neural-functional-training-and-validation-systems}}

\textbf{Table S1.} Training and validation systems for the neural
functionals. Density systems use UCCSD Lambda densities (Section S2) and
their WY orbitals (Section S3); they enter \(\mathcal L_G\) and
\(\mathcal L_D\) during training and the density error during
validation. The C and O atomic energies needed for atomization energies
are evaluated on fixed orbitals but do not enter \(\mathcal L_G\) or
\(\mathcal L_D\). The partition follows Ref. {[}6{]}, with O\(^+\) and
Ne\(^+\) added to the density set.

\begin{longtable}[]{@{}
  >{\raggedright\arraybackslash}p{(\columnwidth - 4\tabcolsep) * \real{0.3333}}
  >{\raggedright\arraybackslash}p{(\columnwidth - 4\tabcolsep) * \real{0.3333}}
  >{\raggedright\arraybackslash}p{(\columnwidth - 4\tabcolsep) * \real{0.3333}}@{}}
\toprule\noalign{}
\begin{minipage}[b]{\linewidth}\raggedright
\end{minipage} & \begin{minipage}[b]{\linewidth}\raggedright
Training
\end{minipage} & \begin{minipage}[b]{\linewidth}\raggedright
Validation
\end{minipage} \\
\midrule\noalign{}
\endhead
\bottomrule\noalign{}
\endlastfoot
Atomization & H\(_2\), LiH, O\(_2\), CO & N\(_2\), NO, F\(_2\), HF \\
Density & H, Li, Ne, H\(_2\), Li\(_2\), LiH, B\(_2\), O\(_2\), CO,
O\(^+\), Ne\(^+\) & He, Be, N, N\(_2\), F\(_2\), HF \\
Ionization & O, Ne & N, F \\
\end{longtable}

The atomization targets are W4-11 values {[}13{]} and the ionization
targets experimental first ionization energies {[}14{]}.

\hypertarget{s14.2.-molecular-test-sets-and-the-density-metric}{%
\subsubsection{S14.2. Molecular test sets and the density
metric}\label{s14.2.-molecular-test-sets-and-the-density-metric}}

AE103 contains 103 molecules with 2--14 atoms from the first three rows,
and IP18 contains the first ionization energies of H--Ar; both are taken
from the data set of Ref. {[}6{]}. The extrapolation subsets contain the
AE103 molecules with an element absent from training (``new elements'',
37 molecules) and those with a bonded atom pair absent from training
(``new bond pairs'', 53 molecules).
\textcolor{red}{[TO FILL: molecule lists for AE103 and both subsets, and IP18 reference values.]}

For a density record \(m\) with self-consistent density
\(n^{\mathrm{SCF}}_m\) and reference \(n^{\mathrm{ref}}_m\), the density
error is

\[
\mathrm{DP}_m=\int\big[n^{\mathrm{SCF}}_m(\mathbf r)-n^{\mathrm{ref}}_m(\mathbf r)\big]^2d\mathbf r,
\qquad
\mathrm{DP}=\frac{\sum_m\omega_m\mathrm{DP}_m}{\sum_m\omega_m},
\tag{S87}
\]

evaluated on the PySCF level-6 grid and reported as
\(\mathrm{DP}\times10^{3}\) in atomic units. DP99 contains 99 records
for 98 distinct molecules; the duplicated ethylene-oxide record gives
that molecule weight two (\(\omega_m=1\) per record). Its references are
UCCSD Lambda densities in the aug-cc-pVTZ basis (Section S2), with the
corrected CH\(_3\)CN and N\(_2\)H\(_4\) geometries and the \(^2\Sigma\)
reference state of C\(_2\)H. DP228 contains 228 density records disjoint
from DP99, none of which enters any fit.
\textcolor{red}{[TO FILL: DP228 reference method, basis, molecule list, and record weights.]}
\textcolor{red}{[TO FILL: Table of AE103 MAE and DP99 values for all functionals in Fig.~3.]}

\hypertarget{s14.3.-gscdb137-sets-and-the-normalized-error-ratio}{%
\subsubsection{S14.3. GSCDB137 sets and the normalized error
ratio}\label{s14.3.-gscdb137-sets-and-the-normalized-error-ratio}}

For a GSCDB137 data set \(d\), the normalized error ratio of a
functional is {[}10{]}

\[
\mathrm{NER}_d=\frac{\mathrm{MAE}_d}{E^{\mathrm{std}}_d},
\qquad
\overline{\mathrm{NER}}=\frac1{|\mathcal S|}\sum_{d\in\mathcal S}\mathrm{NER}_d,
\tag{S88}
\]

where \(E^{\mathrm{std}}_d\) is the mean of the second- to fourth-lowest
errors of the hybrid functionals assessed in Ref. {[}10{]}. The mean in
Eq. (S88) gives every data set equal weight. NER \(<1\) means an error
below that of a typical leading hybrid.

\textbf{Table S2.} GSCDB137 training sets for CDF26D (25 sets, 1500
reaction records).

\begin{longtable}[]{@{}lrllrl@{}}
\toprule\noalign{}
Set & Records & Basis & Set & Records & Basis \\
\midrule\noalign{}
\endhead
\bottomrule\noalign{}
\endlastfoot
BDE99nonMR & 83 & def2-QZVPPD & BHROT27 & 27 & def2-QZVPPD \\
SIE4x4 & 16 & def2-QZVPPD & HB49 & 49 & def2-QZVPPD \\
G21IP & 36 & aug-pc-4 & CHB6 & 6 & def2-QZVPPD \\
DBH22 & 22 & def2-QZVPPD & TA13 & 13 & def2-QZVPPD \\
ISO34 & 34 & def2-QZVPPD & ACONF & 15 & def2-QZVPPD \\
A24 & 24 & def2-QZVPPD & TAUT15 & 15 & def2-QZVPPD \\
AHB21 & 21 & def2-QZVPPD & SN13 & 13 & def2-QZVPPD \\
RG18 & 18 & def2-QZVPPD & ISOMERIZATION20 & 20 & def2-QZVPPD \\
TMD10 & 10 & def2-QZVPP & NC11 & 11 & def2-QZVPPD \\
TAE\_W4-17nonMR & 183 & def2-QZVPPD & G2RC24 & 24 & def2-QZVPPD \\
HAT707nonMR & 505 & def2-QZVPPD & A19Rel6 & 114 & def2-QZVPPD \\
BH46 & 46 & def2-QZVPPD & MB08-165 & 165 & def2-QZVPPD \\
BH76RC & 30 & def2-QZVPPD & & & \\
\end{longtable}

\textbf{Table S3.} GSCDB137 sets not used in fitting CDF26D. Test: 8
energy sets (254 records). EF: electric-field sets.

\begin{longtable}[]{@{}llrl@{}}
\toprule\noalign{}
Group & Set & Records & Basis \\
\midrule\noalign{}
\endhead
\bottomrule\noalign{}
\endlastfoot
Test & G21EA & 25 & aug-pc-4 \\
Test & IP23 & 23 & aug-pc-4 \\
Test & CT20 & 20 & def2-QZVPPD \\
Test & HCP32 & 32 & def2-QZVPPD \\
Test & WCPT26 & 26 & def2-QZVPPD \\
Test & EA50 & 50 & aug-pc-4 \\
Test & ORBH35 & 35 & def2-QZVPPD \\
Test & RSE43 & 43 & def2-QZVPPD \\
EF & Dip146, HR46, Pol130, OEEF & & as in Ref. {[}10{]} \\
\end{longtable}

Record counts follow the protocol manifest, not the nominal numbers in
the set names. The training/test assignment is by participation in the
fit: no test set enters the XC fit (S13.4) or the dispersion fit
(S13.7).

\hypertarget{s15.-additional-results-for-the-neural-functionals}{%
\subsection{S15. Additional results for the neural
functionals}\label{s15.-additional-results-for-the-neural-functionals}}

\begin{figure}
\centering
\includegraphics[width=1\textwidth,height=\textheight]{../figures/fig2_errors.png}
\caption{Absolute errors of self-consistent atomization and ionization
energies of the neural functionals on the test sets. Boxes:
interquartile range; bars: medians; diamonds: mean absolute errors
(MAEs). ``Energy loss only'' denotes E training; the other NN
functionals use the full objective (EGD). CCSD uses the aug-cc-pVTZ
basis.}
\end{figure}

\emph{Energies.} For NN-GGA, EGD lowers the AE103 MAE from 11.6 (E) to
5.6 kcal/mol (PBE: 14.6 kcal/mol). For NN-mGGA, EGD lowers it from 7.1
to 3.3 kcal/mol, below SCAN (3.7 kcal/mol). The gain is largest outside
the training distribution. On new elements, NN-GGA(E) is worse than PBE
(11.2 vs 9.1 kcal/mol): the fixed-orbital error reduced in the fit does
not carry over to self-consistent prediction, as Eq. (3) of the main
text predicts. EGD lowers this MAE to 6.2 kcal/mol, and that of NN-mGGA
from 5.1 to 4.4 kcal/mol. On new bond pairs, EGD lowers the NN-GGA MAE
from 12.0 to 5.0 kcal/mol (PBE: 17.5 kcal/mol) and the NN-mGGA MAE from
9.1 to 2.5 kcal/mol, a factor of 3.6. On IP18, EGD removes the
degradation caused by energy-only training (NN-GGA: 4.3 to 3.2 kcal/mol;
PBE: 3.6 kcal/mol).

\emph{Energies and densities together.} Both E models have larger DP99
errors than their base functionals (NN-GGA(E) 2.3 vs PBE 1.8; NN-mGGA(E)
0.75 vs SCAN 0.64). For NN-GGA the degradation accompanies an energy
gain, reproducing under controlled conditions the ``better energies,
worse densities'' trend of Ref. {[}15{]}. EGD reverses it: density
errors are 2.4 (GGA) and 2.6 (meta-GGA) times lower than with E, and 1.9
and 2.2 times lower than PBE and SCAN. NN-GGA(EGD) has the lowest
density error (0.94) of all GGAs examined and beats 24 of the 26
conventional GGAs on both objectives. TH4, the other GGA on the Pareto
front, and TH3 were both fitted to ab initio XC potentials {[}16{]}; EGD
reaches the same end of the front from six energies and 11 densities.
Among meta-GGAs, NN-mGGA(EGD) has an AE103 MAE of 3.3 kcal/mol (SCAN
3.7, r\(^2\)SCAN {[}18{]} 3.8) and a density error of 0.29, about half
that of SCAN and r\(^2\)SCAN (0.57--0.64) and the lowest of all
meta-GGAs examined. It beats 18 of the 19 conventional meta-GGAs on both
objectives, the exception being VSXC {[}17{]} (3.0 kcal/mol, with a 4.2
times larger density error). SCAN satisfies all 17 exact constraints
known for a meta-GGA {[}5{]}; six energies and 11 densities suffice for
EGD to improve on it in both energy and density.

\hypertarget{references}{%
\subsection{References}\label{references}}

{[}1{]} R. J. Bartlett and M. Musiał, ``Coupled-cluster theory in
quantum chemistry,'' \emph{Reviews of Modern Physics} \textbf{79},
291--352 (2007).
\href{https://doi.org/10.1103/RevModPhys.79.291}{doi:10.1103/RevModPhys.79.291}.

{[}2{]} Q. Wu and W. Yang, ``A direct optimization method for
calculating density functionals and exchange--correlation potentials
from electron densities,'' \emph{The Journal of Chemical Physics}
\textbf{118}, 2498--2509 (2003).
\href{https://doi.org/10.1063/1.1535422}{doi:10.1063/1.1535422}.

{[}3{]} T. Heaton-Burgess, F. A. Bulat, and W. Yang, ``Optimized
Effective Potentials in Finite Basis Sets,'' \emph{Physical Review
Letters} \textbf{98}, 256401 (2007).
\href{https://doi.org/10.1103/PhysRevLett.98.256401}{doi:10.1103/PhysRevLett.98.256401}.

{[}4{]} J. P. Perdew, K. Burke, and M. Ernzerhof, ``Generalized Gradient
Approximation Made Simple,'' \emph{Physical Review Letters} \textbf{77},
3865--3868 (1996).
\href{https://doi.org/10.1103/PhysRevLett.77.3865}{doi:10.1103/PhysRevLett.77.3865}.

{[}5{]} J. Sun, A. Ruzsinszky, and J. P. Perdew, ``Strongly Constrained
and Appropriately Normed Semilocal Density Functional,'' \emph{Physical
Review Letters} \textbf{115}, 036402 (2015).
\href{https://doi.org/10.1103/PhysRevLett.115.036402}{doi:10.1103/PhysRevLett.115.036402}.

{[}6{]} M. F. Kasim and S. M. Vinko, ``Learning the Exchange-Correlation
Functional from Nature with Fully Differentiable Density Functional
Theory,'' \emph{Physical Review Letters} \textbf{127}, 126403 (2021).
\href{https://doi.org/10.1103/PhysRevLett.127.126403}{doi:10.1103/PhysRevLett.127.126403}.

{[}7{]} Q. Sun \emph{et al.}, ``Recent developments in the PySCF program
package,'' \emph{The Journal of Chemical Physics} \textbf{153}, 024109
(2020). \href{https://doi.org/10.1063/5.0006074}{doi:10.1063/5.0006074}.
The numerical implementation described here was inspected against the
installed PySCF 2.14.0 source.

{[}8{]} PySCF source documentation for
\href{https://pyscf.org/_modules/pyscf/cc/uccsd_rdm.html}{UCCSD reduced
density matrices} and
\href{https://pyscf.org/_modules/pyscf/cc/gccsd_rdm.html}{spin-orbital
CCSD reduced density matrices}, in particular
\nolinkurl{_gamma1_intermediates}, \nolinkurl{_make_rdm1}, and
\nolinkurl{make_rdm1}. These routines specify the density-index
convention, Hermitization, reference-occupation addition, and exclusion
of orbital-response contributions used in Section S2.

{[}9{]} H. S. Yu, X. He, S. L. Li, and D. G. Truhlar, ``MN15: A
Kohn--Sham global-hybrid exchange--correlation density functional with
broad accuracy for multi-reference and single-reference systems and
noncovalent interactions,'' \emph{Chemical Science} \textbf{7},
5032--5051 (2016).
\href{https://doi.org/10.1039/C6SC00705H}{doi:10.1039/C6SC00705H}.

{[}10{]} J. Liang and M. Head-Gordon, ``Gold-Standard Chemical Database
137 (GSCDB137): A diverse set of accurate energy differences for
assessing and developing density functionals,'' \emph{Journal of
Chemical Theory and Computation} (2025).
\href{https://doi.org/10.1021/acs.jctc.5c01380}{doi:10.1021/acs.jctc.5c01380}.

{[}11{]} Y. Liu, C. Zhang, Z. Liu, D. G. Truhlar, Y. Wang, and X. He,
``Supervised learning of a chemistry functional with damped
dispersion,'' \emph{Nature Computational Science} \textbf{3}, 48--58
(2023).
\href{https://doi.org/10.1038/s43588-022-00371-5}{doi:10.1038/s43588-022-00371-5}.

{[}12{]} E. Caldeweyher, S. Ehlert, A. Hansen, H. Neugebauer, S.
Spicher, C. Bannwarth, and S. Grimme, ``A generally applicable
atomic-charge dependent London dispersion correction,'' \emph{The
Journal of Chemical Physics} \textbf{150}, 154122 (2019).
\href{https://doi.org/10.1063/1.5090222}{doi:10.1063/1.5090222}.

{[}13{]} A. Karton, S. Daon, and J. M. L. Martin, ``W4-11: A
high-confidence benchmark dataset for computational thermochemistry
derived from first-principles W4 data,'' \emph{Chemical Physics Letters}
\textbf{510}, 165--178 (2011).
\href{https://doi.org/10.1016/j.cplett.2011.05.007}{doi:10.1016/j.cplett.2011.05.007}.

{[}14{]} A. Kramida, Yu. Ralchenko, J. Reader, and NIST ASD Team,
\emph{NIST Atomic Spectra Database}, National Institute of Standards and
Technology. \url{https://physics.nist.gov/asd}.

{[}15{]} M. G. Medvedev, I. S. Bushmarinov, J. Sun, J. P. Perdew, and K.
A. Lyssenko, ``Density functional theory is straying from the path
toward the exact functional,'' \emph{Science} \textbf{355}, 49--52
(2017).
\href{https://doi.org/10.1126/science.aah5975}{doi:10.1126/science.aah5975}.

{[}16{]} N. C. Handy and D. J. Tozer, ``The development of new
exchange-correlation functionals: 3,'' \emph{Molecular Physics}
\textbf{94}, 707--715 (1998).
\href{https://doi.org/10.1080/002689798167863}{doi:10.1080/002689798167863}.

{[}17{]} T. Van Voorhis and G. E. Scuseria, ``A novel form for the
exchange-correlation energy functional,'' \emph{The Journal of Chemical
Physics} \textbf{109}, 400--410 (1998).
\href{https://doi.org/10.1063/1.476577}{doi:10.1063/1.476577}.

{[}18{]} J. W. Furness, A. D. Kaplan, J. Ning, J. P. Perdew, and J. Sun,
``Accurate and numerically efficient r\(^2\)SCAN meta-generalized
gradient approximation,'' \emph{The Journal of Physical Chemistry
Letters} \textbf{11}, 8208--8215 (2020).
\href{https://doi.org/10.1021/acs.jpclett.0c02405}{doi:10.1021/acs.jpclett.0c02405}.